\documentclass[aps,prd,twocolumn,superscriptaddress,groupedaddress,preprintnumbers,nofootinbib]{revtex4-1}  
\usepackage{graphicx}
\usepackage{bm}
\usepackage{epsf,epsfig}
\usepackage{amscd}
\usepackage{amsmath}
\usepackage{amssymb}
\usepackage{hyperref}
\hypersetup{colorlinks=true,linkcolor=blue,citecolor=red,urlcolor=blue}

\newcommand{\nn}{\nonumber\\}

\def\ord#1{{\cal O}(#1)}

\def\CD{\mathcal{D}}

\def\CH{\mathcal{H}}

\def\CO{\mathcal{O}}

\def\CU{\mathcal{U}}

\def\p{\partial}
\def\d{{\rm d}}

\begin{document}

\title{Inflationary Potentials from the Exact Renormalisation Group}
\author{Sa\v{s}o Grozdanov} 
\email{grozdanov@lorentz.leidenuniv.nl}
\affiliation{Instituut-Lorentz for Theoretical Physics, Leiden University,\\Niels Bohrweg 2, Leiden 2333 CA, The Netherlands}
\author{David Kralji\'{c}}
\email{david.kraljic@physics.ox.ac.uk}
\affiliation{Rudolf Peierls Centre for Theoretical Physics, University of Oxford, \\1 Keble Road, Oxford OX1 3NP, U.K.}
\author{Eirik Eik Svanes}
\email{esvanes@lpthe.jussieu.fr}
\affiliation{LPTHE, UMR 7589, Sorbonne Universit\'es, UPMC Paris 06, F-75005, Paris, France \\ CNRS, UMR 7589, LPTHE, F-75005, Paris, France\\  Institut Lagrange de Paris, Sorbonne Universit\'es, 98 bis Bd Arago, 75014 Paris, France}

\begin{abstract}
We show that an inflationary slow-roll potential can be derived as an IR limit of the non-perturbative exact renormalisation group equation for a scalar field within the mean-field approximation. The result follows without having to specify a Lagrangian in the UV, which we take to be somewhere below the Planck scale to avoid discussing quantum gravity effects. We assume that the theory contains a scalar mode with suppressed coupling to other fields, and that higher derivative couplings are suppressed. In this framework the exact RG equation becomes a one-dimensional Schr\"odinger equation, which we solve. The effective IR potential is then dominated by the eigen-states of the RG Hamiltonian with the highest eigenvalues. We find that these potentials can generically give rise to slow-roll inflation, which is fully consistent with recent observations. As an example of how the proposed renormalisation group procedure works, we perform an explicit calculation in the $\phi^4$ theory in an appendix.
\end{abstract}

\maketitle

\begingroup
\hypersetup{linkcolor=black}
\tableofcontents
\endgroup

\section{Introduction}
Inflation \cite{PhysRevD.23.347,starinf,Linde:1981mu} is an exponentially fast expansion in the early universe that is claimed to resolve a number of problems of the standard Big Bang cosmology (e.g. the horizon, flatness, smoothness, relic problems). It also provides a mechanism for generation of the density perturbations \cite{mukhfluct} that have left an imprint on the cosmic microwave background and grown into the observed large-scale structure of the universe \cite{inflbook}. Models of inflation are typically discussed within the slow-roll paradigm, where an inflaton field evolves along a nearly flat potential to its minimum, during which it sources a quasi-exponential growth of the cosmological scale factor. Hence, in the slow-roll regime, the spatial and time derivatives are negligible compared to the approximately constant vacuum energy. 

Many potentials driving the inflationary fields have been proposed that are either phenomenological or inspired by various UV theories \citep{encyclinf}. In this paper, we propose a different approach to deriving a scalar field theory suitable for the slow-roll inflation: We will argue that inflation can be understood as a very generic, non-perturbative prediction of the exact renormalisation group (ERG), in a way that is largely insensitive to the details of the UV physics which we take to be at a scale somewhere below the Planck scale, $M_{UV}<M_{Pl}$, in order to ignore quantum gravity effects. We add that the observation that renormalisation group dynamics can produce experimentally viable inflation scenarios has been noted in the literature before \cite{Salvio:2014soa, Kannike:2015apa}. Moreover, exact RG has also been applied in the context of inflation before, see e.g. \cite{Copeland:2013vva, Kaya:2013bga, Saltas:2015vsc}. 

To have a valid potential for the description of inflation, we must ensure that the effective theory is valid below some scale $M$, where $M$ is roughly on the order of the inverse Hubble radius at the start of inflation. This is the IR of our theory. Thus, $1/M^4$ will be approximately the volume of the patch of spacetime of the resulting effective theory, which we assume resides within the Hubble radius. We will assume that the UV theory at $M_{UV}>>M$ contains a scalar mode $\phi$ that couples very weakly to other fields. Hence, during the RG flow from $M_{UV}$ to $M$, the corrections to the effective potential of such a scalar from the couplings to other UV fields would remain small. In the RG analysis, $\phi$ can therefore be treated independently. 

The second assumption will be the validity of the mean-field approximation (MFA) near the scale $M$.\footnote{For other recent implementations of the ERG within the MFA, in particular in de Sitter space, see \cite{serreau2014renormalization, benedetti2015critical, Guilleux:2015pma}.} That is, we will treat the constant mode of the theory separately when we do our IR analysis. In the paper, we will give an argument for the plausibility of this assumption, given that a field can be expanded around its constant value. We work in the local potential approximation, where higher derivative couplings are ignored. The argument will then rely on the particular form of the ERG equation. As a result, we will only be interested in the constant mode of the theory, which will be sufficient to describe the effective potential. Without justification, this assumption may at first seem peculiar, but we note that for inflation to begin, the inflaton field has to be sufficiently smooth to overcome the gradient energy preventing the exponential expansion. Indeed, the MFA is a common feature of many other inflationary scenarios, where the field is assumed constant at the scale where inflation begins \cite{causalVachas}. 

Lastly, we will also ignore effects of gravity in the ERG. We start at a scale  $M_{UV}$ somewhere below the Planck scale, so that quantum gravity effects can be ignored. For the most part of the RG flow, we are at an energy scale where the vacuum energy is negligible in comparison, and we can work in Minkowski space where gravitational effects are unimportant. Indeed, the de Sitter value of the vacuum energy of the universe is roughly comparable with the scale at which inflation begins, which is where we end our flow. Flows beyond this point would require us to take gravitational effects into account \cite{serreau2014renormalization, benedetti2015critical, Guilleux:2015pma}.

After introducing the ERG equation relevant for our setup and justifying the validity of the MFA, we will show that the MFA of the ERG equation gives rise to a one-dimensional Schr\"odinger-type equation for the resulting effective theory, which can be solved exactly. This is similar to the stochastic approach of \cite{starobinsky1986stochastic, Starobinsky:1994bd} as well as the study of renormalisation group in quantum mechanics \cite{Polonyi:1994pn}. In essence, we will solve a simple quantum mechanical problem in which the resulting ``wave-function" will correspond to the effective theory, or potential, for the constant mode. Because in QFT the theories need not be normalisable functions of the field variable, we will only insist that the resulting potential is bounded from below. What is meant by normalisable and un-normalisable solutions will become clear below. Our analysis will then result in non-perturbative IR effective potentials with shapes suitable for inflation. It will give predictions that are fully consistent with current observations. We note that spectral methods have been applied in the context of ERG before, see e.g. \cite{Litim:2003kf, Zappala:2012wh, Fischer:2004uk, Borchardt:2015rxa, Borchardt:2016pif} and references therein for more details.

At the end of the paper, we provide two appendices: \ref{app:completeset} is devoted to an ERG analysis of scalar field theories, in particular the $\phi^4$-theory, which is an example of a normalisable theory. \ref{app:Momspace} is devoted to solving the Schr\"odinger equation in momentum space, where the solution takes a rather simple form. 

\section{Exact renormalisation group analysis}\label{Ref:ERG}
\subsection{The ERG equation}
Let us begin by considering a UV quantum field theory of the early universe that contains some set of fields at the UV scale $M_{UV}<M_{Pl}$, including a scalar mode $\phi$. By $\Phi$, we collectively denotes the remaining UV fields. As the theory runs between $M_{UV}$ and the IR scale $M$, we assume that $\phi$ and $\Phi$ interact very weakly so that the IR theory for $p < M$ takes the form,
\begin{align}\label{PartitionFunction}
Z =\int_M \CD \phi \,  e^{- S_\phi \left[\phi\right]}  \int_M \CD \Phi \, e^{ - S_{\Phi} \left[\Phi \right] }.
\end{align} 
We can thus focus only on the decoupled renormalisation group flow for the scalar $\phi$ by following the procedure of the exact Wilsonian renormalisation group \cite{Wilson:1973jj}.\footnote{For reviews on the exact renormalisation group, see \cite{Polonyi:2001se,Bervillier:2013kda,Bagnuls:2000ae}.} The ERG equation for the decoupled scalar mode takes the form
\begin{align}\label{ERGEp}
\p_t S_\phi = \!\! \int_p \! \left(\alpha(t)+2p^2\right)  \! \left[\frac{\delta^2S_\phi}{\delta\phi_p\delta\phi_{-p}}-\frac{\delta S_\phi}{\delta\phi_p}\frac{\delta S_\phi}{\delta\phi_{-p}}+\phi_p\frac{\delta S_\phi}{\delta\phi_p}\right],
\end{align}
where $\alpha(t)$ depends on the choice of the cut-off and $t \sim 1/ \Lambda$ denotes the RG time -- an inverse of the cut-off scale $\Lambda$.\footnote{Since we are only interested in solving the theory in the IR, we do not include the rescaling of the theory back to the original UV scale -- blocking from $M_{UV}$ to $M$ suffices.} From an inflationary standpoint it is also convenient to measure the dimensionless field $\phi$ in units of $M_{Pl}$, rather than $M_{UV}$.\footnote{Throughout this paper, we use the reduced Planck mass, $M_{Pl}=2.44 \times 10^{18}$\,GeV.} 

The ERG equation \eqref{ERGEp} is a non-linear equation. However, if we instead consider the functional 
\begin{align}
\Psi\left[\phi\right]= e^{-S_\phi \left[\phi\right]},
\end{align}
then Eq. \eqref{ERGEp} becomes a linear Schr\"{o}dinger-type equation, 
\begin{align}\label{eq:SchrodingerRG}
\p_t\Psi \left[t,\phi\right] =\hat\CH \Psi\left[t,\phi\right],
\end{align}
with a Hamiltonian operator in position space,
\begin{align}\label{SchrodingerHam}
\hat\CH = \int d^4 x\left(\alpha(t)-2\p^2\right)\left(\frac{\delta^2}{\delta\phi_x^2 }+\phi_x \frac{\delta}{\delta\phi_x}\right).
\end{align}
This is the usual form of the Wilsonian ERG equation, written in terms of $\Psi$. Similar equations have been derived in the literature, see e.g. \cite{polchinski1984renormalization, comellas1997n}. Note that one common feature of these equations is that the Hamiltonian has a kinetic part $\delta^2/\delta\phi^2$, together with a divergence term $\phi\, \delta/\delta\phi$.

\subsection{Validity of the MFA}
\label{sec:MFA}
Before performing a detailed analysis of Eq. \eqref{eq:SchrodingerRG}, we pause to discuss more precisely why the MFA is a valid approximation of the ERG equation when we are only interested in the potential of the theory for an approximately constant field. Working in the local potential approximation, we find that the kinetic modes decouple faster as we move towards the IR, and we integrate them out from the theory. This leaves us with an effective potential for the constant mode. We should also note that our conventions are such that all fields and coordinates will be dimensionless throughout the RG analysis. 

We begin by expanding the functional $\Psi$ as 
\begin{align}
\Psi\left[\phi\right]=&\,\Psi^0\left[\phi_0\right]+\int_{p}\phi_p\Psi^1_p\left[\phi_0\right] \nn
&+\frac{1}{2}\int_{\{p,q\}}\phi_p\phi_q\Psi^2_{p,q}\left[\phi_0\right]+\ldots,
\label{eq:expandPsi}
\end{align}
where $\Psi^1_0=\Psi^2_{0q}=0$, and $\Psi^2_{p,q}$ can be assumed to be symmetric in $p$ and $q$. We ignore terms that are cubic or higher-order in momenta. This is closely related to the often used local potential approximation. From hereon, we will denote the constant mode as $\phi_0=x$. Moreover, 
\begin{equation}
\Psi^0[x]=e^{-S_0(x)},
\end{equation}
which gives the potential of the theory. Inserting this into Eq. \eqref{eq:SchrodingerRG}, we can derive the following set of equations,
\begin{align}
\p_t\Psi^0&=\hat \CH_0 \Psi^0+\int_q\left(\alpha(t)+2q^2\right)\Psi^2_{q,-q},\label{TayExp1}\\
\p_t\Psi^1_p&=\hat \CH_0 \Psi^1_p+\left(\alpha(t)+2p^2\right)\Psi^1_p , \label{TayExp2} \\
\p_t\Psi^2_{p,q}&=\hat \CH_0\Psi^2_{p,q}+2\left(\alpha(t)+p^2+q^2\right)\Psi^2_{p,q}, \label{TayExp3} 
\end{align}
where 
\begin{equation}
\hat \CH_0 =\alpha(t)\left(\p_{x}^2+x\p_{x}\right).
\end{equation}
Eqs. \eqref{TayExp1} and \eqref{TayExp3} can then be put in the form
\begin{equation}
\label{eq:approxERG}
\p_t\left( \begin{array}{c}
\Psi^0 \\
\Psi^2_{p,q} \end{array} \right)=
\left( \begin{array}{cc}
\hat \CH_0 & \tilde{\rm tr} \\
0 & \hat \CH_{p,q} \end{array} \right)
\left( \begin{array}{c}
\Psi^0 \\
\Psi^2_{p,q} \end{array} \right),
\end{equation}
where
\begin{equation}
\label{eq:momentumHam}
\hat \CH_{p,q}=\hat \CH_0+2\left(\alpha(t)+p^2+q^2\right),
\end{equation}
while the trace $\tilde{\rm tr}$ is given by
\begin{equation}
\tilde{\rm tr}(\Psi^2_{p,q})=\int_q \left(\alpha(t)+2q^2\right)\Psi^2_{q,-q}.
\end{equation}
As $t\rightarrow\infty$, we can assume that $\alpha(t)\rightarrow\alpha$ becomes constant \cite{Bagnuls:2000ae}. The solution of \eqref{eq:approxERG}, as $t\rightarrow\infty$, is therefore given by the highest eigenvalue eigenmodes of the matrix
\begin{equation}
\mathbb{H}=
\left( \begin{array}{cc}
\hat \CH_0 & \tilde{\rm tr} \\
0 & \hat \CH_{p,q} \end{array} \right).
\end{equation}
This matrix is upper-triangular, so the eigenmodes factor into the eigenmodes of $\hat \CH_0$ and $\hat \CH_{p.q}$. In particular, the dominant term in the potential will be given by the highest eigenmode of $\hat \CH_0$. 

For completeness, let us also consider what happens to the kinetic terms in the expansion \eqref{eq:expandPsi} as $t\rightarrow\infty$. Note first that the term linear in $\phi_p$, in Eq. \eqref{eq:expandPsi}, integrates to zero, and we will hence ignore it. Note also that in order for eigenmodes of $\hat \CH_0$ not to blow up at infinity, we will see below that the eigenvalue is required to be less than or equal to zero. Finally, note form \eqref{eq:momentumHam} that the kinetic terms $\Psi^2_{p,q}$ will blow up quicker than $\Psi^0$, and will hence integrate out sooner. This is true even for momenta close to zero due to the non-vanishing $\alpha$. Ignoring higher-order kinetic terms, the action reads
\begin{equation}
S[\phi]=-\log(\Psi^0[x])-\frac{1}{2}\int_{\{p,q\}}\phi_p\phi_q\tilde\Psi^2_{p,q}\left[x\right]+\ldots,
\end{equation}
where $\tilde\Psi^2_{p,q}\left[x\right]=\Psi^2_{p,q}\left[x\right]/\Psi^0[x]$. Note that in order to have a positive-definite kinetic term, we require $\Psi^2_{p,q}\left[\phi_0\right]$ to be negative. If we also assume that $\Psi^2_{p,q}$ saturates at the highest allowed mode, which is constant, i.e.
\begin{equation}
\Psi^2_{p,q}\left[\phi_0\right]=-C_{p,q},
\end{equation}
then we can integrate out the momentum modes to get the effective action
\begin{equation}
S(x)=S_0(x)+\frac{1}{2}\int_q\left(S_0(x)+\log(C_{q,q})\right).
\end{equation}
We thus see that the correction only ends up multiplying the potential with an overall constant and adding an irrelevant constant to the potential. Beyond that, the potential remains the same. Hence, we see that in the local potential approximation, we are justified in using the MFA and we will therefore ignore propagating modes for the remainder of the paper.

\subsection{Solution of the RG flow}
We can now use the MFA to solve the ERG equation \eqref{eq:SchrodingerRG} for the patch of the universe of the size of $1/M^4$, i.e. in the IR regime of the theory. In the UV, we wish to remain as general as possible, so we do not specify the details of the theory. Let us then write the Euclidean partition function \eqref{PartitionFunction} with its initial theory specified in the UV as
\begin{align}\label{UVPathIntegral}
Z = \int \CD \phi \, \Psi_{UV} \left[t,\phi\right].
\end{align}
In the MFA, an eigenmode of the Hamiltonian $\hat\CH$ evolves under the ERG equation \eqref{eq:SchrodingerRG} as
\begin{align}
\Psi(t,x) = e^{E t} \Psi(x),
\end{align}
where $E$ are the eigenvalues of the RG time-independent equation
\begin{align}\label{MeanRG2}
\hat\CH \Psi = E \Psi.
\end{align}
Note that we are using $\Psi \equiv \Psi(x)$. In analogy with quantum mechanics, we assume that $\Psi_{UV}(t,x)$ can be expanded as
\begin{equation}\label{CompleteSetExp}
\Psi_{UV}(t,x)=\sum_i \gamma_i e^{E_i t} \Psi_i(x),
\end{equation}
where $\Psi_i$ have eigenvalues $E_i$. In the IR, where $t\to\infty$, the dominant contribution will therefore come from the highest eigenvalue solution, $\Psi_i$ with $\max\left[E_i\right]$, that has a non-trivial overlap with the UV theory. The contribution of other eigenvalue solutions will decay exponentially fast with RG time so we expect the highest eigenvalue solution to dominate even with little RG flow. Note that in Eq. \eqref{CompleteSetExp}, a functional integral over $\Psi_i$ should be used instead of the sum when $E_i$ are a continuous set. 

In Appendix \ref{app:completeset}, we show an example of how the decomposition in Eq. \eqref{CompleteSetExp} can be performed in the mean-field approximation for a UV theory with a ``normalisable", finite (path) integral of $\Psi_{UV}(x)$ over $x$. We use the simplest example of an interacting field theory: the $\phi^4$ theory. In general, however, such an explicit computation of $\gamma_i$ may be extremely challenging, especially for potentials which give formally divergent integrals. Hereon, we will therefore only assume that such a decomposition is possible and that we can treat the most general solution for the ``ground state" theory (one with $\max\left[E_i\right]$) as the dominant theory in the IR. See Appendix \ref{app:completeset} for definitions and further details.

It is convenient to introduce a new functional $\psi$ so that 
\begin{align}
\Psi \equiv e^{- \hat x^2 / 4} \psi .
\end{align} 
The ERG eigenvalue equation in the MFA then becomes 
\begin{align}
\label{MeanRG3}
\hat H \psi = E \psi .
\end{align}
In terms of the familiar quantum mechanical notation, we can use $\hat x \equiv\phi/M_{Pl}$ for our initial UV field $\phi$, where $\hat x$ is measured in units of $M_{Pl}$ as is convenient for computations involving inflation. We also let $\hat p \equiv - i \p_{\hat x}$, which gives the Hamiltonian operator $\hat H$:
\begin{equation}\label{Hofpsi}
\hat H= - \left( \hat p^2 + \frac{\hat x^2}{4} + \frac{1}{2} \right),
\end{equation}
where we have set $\alpha=1$ without loss of generality. It is important to note that the solutions to \eqref{MeanRG2} and \eqref{MeanRG3}, i.e. $\Psi_i$ and $\psi_i$, respectively, have the same eigenvalues, $E_i$. In the MFA, the two Hamiltonians are related by
\begin{equation}
\hat \CH=e^{-x^2/4} \hat H e^{x^2/4}=\p_x^2+x\p_x\:.
\end{equation}

The form of $\hat H$ in Eq. \eqref{Hofpsi} is highly reminiscent of the quantum harmonic oscillator in imaginary (Euclidean) time. Indeed, by defining the ladder operators
\begin{align}
\hat a&=\frac{1}{2}\hat x+i\hat p,\\
\hat a^*&=\frac{1}{2}\hat x-i\hat p,
\end{align}
the Hamiltonian takes the form
\begin{equation}\label{Ham}
\hat H=-(\hat a^*\hat a+1).
\end{equation}
Eq. \eqref{Ham}, which describes the renormalisation group evolution of the effective action thus only differs from the usual harmonic oscillator by an overall minus sign and the additive factor of $1$ in place of $1/2$. 

For modes of $\psi$ which tend to zero as $x\rightarrow\pm\infty$, the operator $a^*$ is the adjoint of $a$, $a^*=a^\dagger$. To see this, let $\alpha$ and $\beta$ be modes that tend to zero at infinity and consider
\begin{equation}
(\hat a\alpha,\beta)=\int \!dx\, (\hat a\alpha) \beta^* =\int \!dx\, \alpha (\hat a^*\beta)^*=(\alpha,\hat a^\dagger\beta),
\end{equation}
where we have performed an integration by parts. $\hat H$ is negative definite for such modes, and the highest eigenvalue is the ``vacuum energy",
\begin{align}\label{E0eigenvalue}
E_0\equiv E =-1,
\end{align}
which corresponds to the vacuum of the theory, $\psi_0$,
\begin{align}
\label{eq:vac1}
\hat a\,\psi_0&=0.
\end{align}
The corresponding re-scaled ``original" theory reads
\begin{align}
\Psi_0 (x)&=C_0 e^{-x^2/4}\psi_0(x)\nn
&=C_0 e^{-x^2/2},
\end{align}
which precisely corresponds to the potential of a free theory. Besides the free theory, there exists a mode, $\tilde\psi_0$, with the same eigenvalue \eqref{E0eigenvalue},
\begin{align}
&\tilde\psi_0 (x)=D_0 e^{-x^2/4} \, \text{erfi}\left(\frac{x}{\sqrt{2}}\right),\\
&\tilde \Psi_0 (x)=D_0  e^{-x^2/4}\tilde\psi_0(x)=D_0 e^{-x^2/2} \,\text{erfi} \left(\frac{x}{\sqrt{2}}\right),
\end{align}
were $\text{erfi}(x)$ denotes the imaginary error function. Note that while $\tilde\psi_0(x)$ diverges as $x\rightarrow\pm\infty$, as expected for a un-normalisable mode of the harmonic oscillator, the re-scaled wave-function, $\tilde \Psi_0$, exhibits the correct behaviour and tends to zero at infinity. $\tilde\Psi_0$ leads to a bounded potential, as 
\begin{align}
S_{\phi} \left[\phi\right]= - \log \tilde\Psi_0=\frac{\tilde V_0(\phi)}{M^4} ,
\end{align} 
where
\begin{equation}
\tilde V_0(\phi)=M^4\left[\log \left(D_0 \right)+\frac{1}{2}x^2- \log \left[\text{erfi} \left(\frac{x}{\sqrt{2}}\right)\right] \right] .
\end{equation}

In order to avoid considering theories of arbitrarily high eigenvalues, one reasonable assumption is that the UV theory has {\em no} overlap with modes $\Psi_i$ of which the eigenvalues would be higher than the ``vacuum" energy $E_0$, cf. Eq. \eqref{E0eigenvalue}. In physical terms, this means that we assume that the IR limit of $\Psi_{UV}$ is connected to the free theory -- the state $\Psi_0$. 

In the absence of such a restriction, it is clear that the eigenstates of \eqref{MeanRG3}, and thus also $\Psi$, could have arbitrarily high eigenvalues. However, we want our dominant IR theory, denoted temporarily by $\Psi = \Psi_E$, with some eigenvalue $E$, to tend to zero as $x\rightarrow\pm\infty$, as required by a stability condition that potentials need to be bounded from below. The eigenvalue equation \eqref{MeanRG2} then reads
\begin{align}
\label{eigen1}
\hat\CH\Psi_E&=E\Psi_E\nn
&=\p_x^2\Psi_E+x\p_x\Psi_E \:.
\end{align}
If we multiply this equation by $\Psi_E$ and integrate over $x$, we get after integration by part,
\begin{equation}\label{NegE}
E\Psi_E=-\int_x(\p_x\Psi_E)^2-\frac{1}{2}\int_x\Psi_E^2 .
\end{equation}
The main point is that integration by parts can only be performed for $\Psi_E$ that tend to zero at infinity. Since the integrands on the right-hand-side of \eqref{NegE} are negative-definite, it follows that theories with bounded potentials must have $E<0$. 

In the remainder of this work, we will focus on two classes of stable effective potentials: the IR effective theories with {\em discrete} and {\em continuous} spectra of $E_i < 0$. In the deep IR limit where $t\to\infty$, the existence of a discrete spectrum with integral eigenvalues implies that the state with the highest eigenvalue, $E_0 = -1$, will dominate the first class of theories. In the continuous case, theories with $E_i$ close to $0$ will dominate. 

We can now restate the above results in the following way: By assuming that $\Psi_{UV}$ has a non-zero overlap only with the states of integral eigenvalues $E=-n$ for $n\in\mathbb{N}$, the IR limit of the theory takes the generic form
\begin{equation}\label{SolPsi}
\Psi_0 (x) =e^{-\frac{1}{2}x^2}\left[C_0+D_0\:\text{erfi}\left(\frac{x}{\sqrt{2}} \right)\right] ,
\end{equation}
where $C_0$ and $D_0$ are arbitrary constants. Importantly, the IR effective theory is continuously connected to the free Gaussian IR limit, as is often the case in perturbative RG. 

If we lift the restriction of including the free Gaussian potential in our IR solution, then it becomes more natural to consider the second, continuous class of theories for which the Eq. \eqref{MeanRG3} (or Eq. \eqref{MeanRG2}) gives
\begin{align}\label{SolPsiGeneral}
\Psi (x) =&\, e^{-\frac{1}{2}x^2} \biggr[ C_0\, {}_{1}F_{1} \left( \frac{1+E}{2}; \frac{1}{2}; \frac{x^2}{2} \right) \nn
&+ D_0 \,x \sqrt{\frac{2}{\pi}}  \,{}_{1}F_{1}\left( \frac{2+E}{2}; \frac{3}{2}; \frac{x^2}{2} \right) \biggr],
\end{align}
where $_{1}F_1\left(a;b;x^2/2\right)$ is the confluent hypergeometric function of the first kind. The solution in Eq. \eqref{SolPsiGeneral} is valid for all $E$ and reduces to \eqref{SolPsi} at $E = -1$. 

A simple complete set of orthogonal polynomials (the Hermite polynomials) can be formed from the $\Psi(x)$ functions in \eqref{SolPsiGeneral} by restricting the eigenvalues $E$ to be negative integers, $E \in \{-1,-2,-3,\ldots\}$. In this way, we can form an infinite series of finite and ``normalisable" contributions to the expansion of $\Psi_{UV}$ in \eqref{CompleteSetExp}. Here, we define normalisable to mean that the integral over $x$ of $\Psi(x)$ is finite for some choice of $E$:
\begin{align}\label{DefNormal}
\int_{-\infty}^\infty dx \,\Psi_E(x) = \text{finite}.
\end{align}
As already noted above, we explore this possibility in detail in Appendix \ref{app:completeset} where we use this basis to decompose the ERG flow of normalisable scalar theories, in particular the $\phi^4$ theory. However, in general, we expect that the integrals over $\Psi(x)$ need not converge to give us a physically acceptable effective potentials. In the remainder of this paper, we will study such ``un-normalisable" theories, for which the integrals diverge:
\begin{align}
\int_{-\infty}^\infty dx \,\Psi_E(x) \to \infty.
\end{align}
In particular, the potentials that we will use as candidates for inflation will be of the un-normalisable type. 

It is important to keep in mind that both MFA solutions, \eqref{SolPsi} and \eqref{SolPsiGeneral}, can be non-perturbative in the coupling constants of the original UV theory. However, the dependence of the effective potential on the couplings could only be computed if we had specified the UV theory and somehow computed the relevant overlap integrals that would reveal the true weights of $\Psi_i$ in $\Psi$. We will not pursue this direction in this work and will only consider IR effective potentials with unspecified coupling constant dependence. 

Before we move on to considering the phenomenological implication of the two solutions, we note that our approach resembles that of the stochastic approach discussed in \cite{starobinsky1986stochastic, Starobinsky:1994bd}. It is also similar to the approach of Halpern and Huang \cite{Halpern:1995vf} and Periwal \cite{Periwal:1995hw}, where a linearisation of the ERG equation \eqref{ERGEp} was performed. Those works showed that there are non-polynomial deformations of the free Gaussian theory for which the Gaussian is IR-unstable. Such cases are also included in our analysis, within the MFA. Furthermore, this gives credence to the inclusion of the extra mode $\tilde\psi_0$ in addition to the Gaussian when considering theories of eigenvalue $-1$. It also prompts us to consider in detail theories with eigenvalue greater than $-1$, which can be argued to be preferable over the free theory from the exact renormalisation group point of view. 

Finally, we also note that our non-perturbative result differs from the conventional perturbative approach where the effective Lagrangian is expanded in the powers of the field suppressed by some mass scale. Our approach is also different to what is known as the Effective Field Theory of Inflation \citep{eftinf}, where the background dynamics sourced by a potential is assumed as given and the effective theory refers to the fluctuations around that background.

\section{Inflation}
\subsection{Inflationary potential}
Having found solutions that are likely to dominate the non-perturbative IR regimes of scalar theories, we now turn our attention to studying the phenomenological implications of such theories. In particular, we will show that solutions of Eq. \eqref{eq:SchrodingerRG} naturally lead to potentials capable of sustaining inflation in the early universe. The physical requirement in the patch of spacetime where inflation starts is that the temporal derivative of the field is small and the field is homogeneous \citep{causalVachas,goldinitinf}. To describe inflation, we are therefore primarily interested in the potentials for the constant mode of the theory, which given the argument in section \ref{sec:MFA} means that we can use our MFA results, where we effectively ignored the higher momentum modes in order to derive equation \eqref{MeanRG3} and its solutions.

We begin by restating the un-normalisable solution with a continuous spectrum of $E_i$, i.e. Eq. \eqref{SolPsiGeneral}, as an effective potential of $\phi$. Restoring the mass dimensions,

\begin{widetext}
\begin{align}
&\frac{V(\phi)}{M^4} =  \frac{1}{2} \left(\frac{\phi}{M_{pl}}\right)^2 -\log\left[  _{1}F_{1}\left( \frac{1+E}{2}; \frac{1}{2}; \frac{1}{2} \left(\frac{\phi}{M_{pl}}\right)^2 \right) + c \,\sqrt{\frac{2}{\pi}}\left(\frac{\phi}{M_{pl}}\right) {}_{1}F_{1}\left( \frac{2+E}{2}; \frac{3}{2}; \frac{1}{2} \left(\frac{\phi}{M_{pl}}\right)^2 \right) \right] + \frac{C(c)}{M^4},\label{eq:epspotential}
\end{align}
\end{widetext}

where $M_{Pl}$ is the reduced Planck mass.. The potential has an overall factor expressed in terms of some mass scale $M$. Furthermore, we have two integration constants $c$ and $C$. Our universe has a very small cosmological constant, therefore the constant $C$ needs to be fixed so that the vacuum energy at the minimum of the potential is zero, $V_{min}=0$. This constrains $C$ to be a function of $c$, i.e. $C(c)$.

\begin{figure}[t]
\includegraphics[width=\columnwidth]{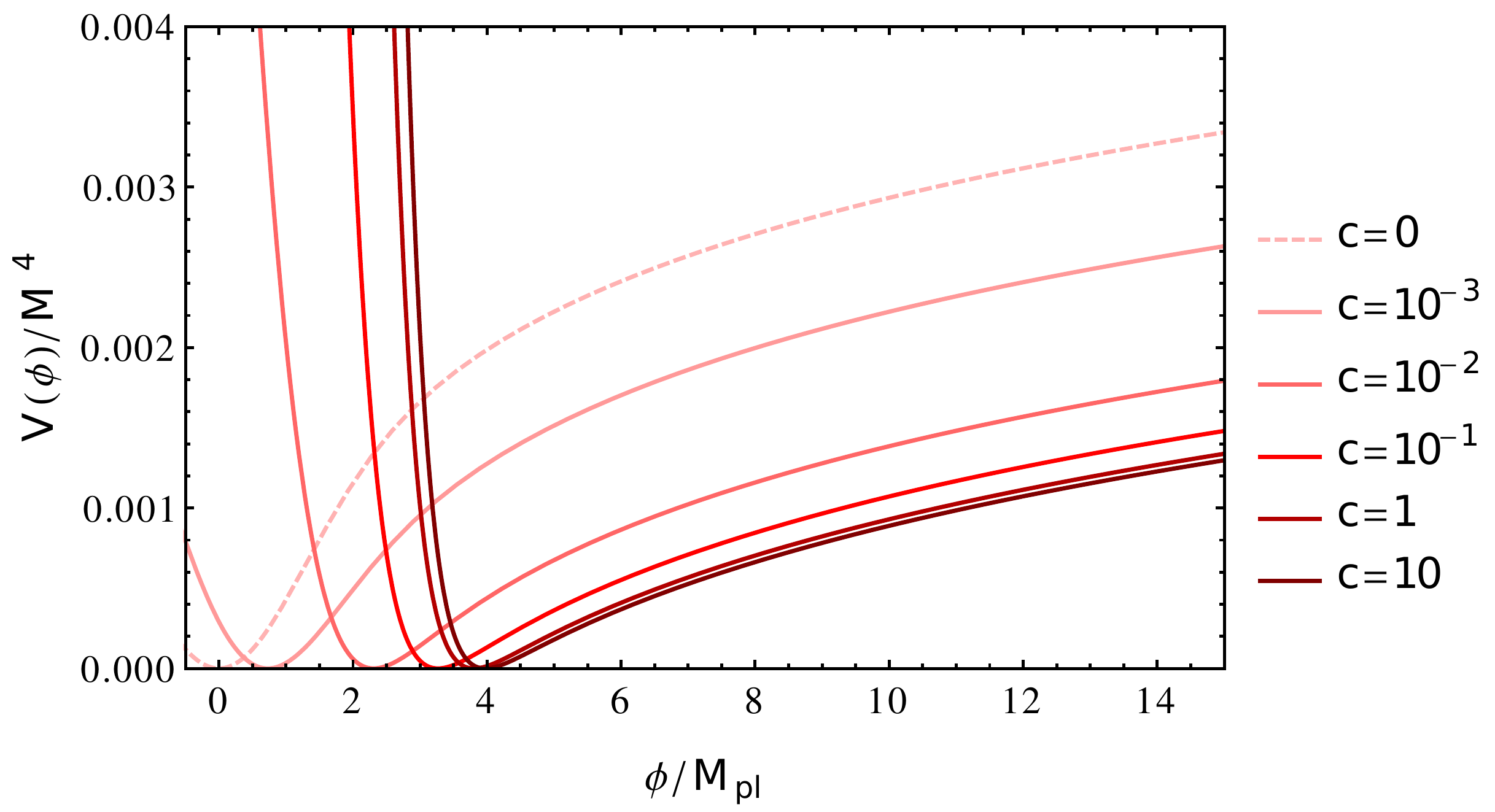}
\caption{The dependence of the potential from Eq. \eqref{eq:epspotential} on the parameter $c$. Note that the shape does not change for $c \gtrsim \ord{1}$. Here, we choose $E\sim-10^{-3}$.}
\label{fig:epspotentials}
\end{figure}

At the minimum of the potential, the mass of the scalar field is independent of the shape parameter $c$,
\begin{align}
\p^2_\phi V (\phi_{min})=- \frac{E M^4}{M_{Pl}^2}.
\end{align}
When $c \gtrsim 1$, the potential and its derivatives do not depend on $c$ (see Fig. \ref{fig:epspotentials}). Hence, the potential has a stable shape and the exact value of $c$ for $c \gtrsim \ord{1}$  does not matter.

We are particularly interested in the cases that dominate the IR regime of the RG flow, where $E\rightarrow0^-$. For small $|E|$, the plateau region of the potential is not only flat, but also small (see Fig. \ref{fig:epspotentials}). Such a potential can support slow-roll inflation as well as result in small amplitude of the scalar perturbations required by the observations, thus alleviating the fine-tuning of $M$ (see next section). 

Let us explore the plateau region. We can use the asymptotic expansion of the hypergeometric functions to obtain the leading behaviour that is logarithmic:
\begin{equation}
V(\phi)\approx M'^4\left[ 1 + \alpha \log\left(\frac{\phi}{M_{Pl}}\right) \right]. \label{eq:logarithmic}
\end{equation}
This potential shape is reminiscent of the `Loop Inflation' model\footnote{Frequently termed `Spontaneously broken SUSY' model.} where the logarithmic dependence arises from the loop corrections that ``spoil'' the flatness of the inflationary potential (i.e. the $\eta$-problem). This has been studied in the context of the $F$- and $D$-term inflation.\footnote{For a review of these approaches, see Ref. \citep{encyclinf} and references therein.} 

\subsection{The slow-roll analysis in theories with a continuous spectrum of $E$}
The background dynamics of the homogeneous scalar field in the FRW universe is governed by the continuity and Friedmann equations:
\begin{align}
&\ddot{\phi}+3H\dot{\phi}+\p_\phi V =0 ,\\
&H^2 = \frac{1}{3M_{Pl}^2}\left( \frac{1}{2}\dot{\phi}^2 + V(\phi) \right).
\end{align}
Here, $H=\dot{a}/a$ is the Hubble parameter and $a(t)$ is the scale factor that depends on time, not the RG time that we used in Sec. \ref{Ref:ERG}. It follows then that the accelerated expansion of the universe ($\ddot{a}/a>0$) is achieved when $\dot{\phi}^2<V(\phi)$, that is, the potential energy of the scalar field dominates over the kinetic energy. Sustaining the accelerated expansion for long enough also requires the second derivative of the field $\ddot{\phi}$ to be small. These conditions can be encoded in the smallness of two potential dependent slow-roll parameters:
\begin{equation}
\label{eq:slowroll}
\begin{aligned}
&\epsilon_V(\phi) \equiv \frac{M_{Pl}^2}{2}\left(\frac{\p_\phi V }{V}\right)^2,\\
&\eta_V(\phi) \equiv M_{Pl}^2 \left(\frac{\p^2_\phi V}{V}\right). 
\end{aligned}
\end{equation}
Inflation proceeds when $\epsilon_V\ll1$ and $|\eta_V| \ll 1$, and ends when $\epsilon_V(\phi_{end})\approx 1$. During the expansion, the universe grows by a number of e-folds:
\begin{equation}
N(\phi)\equiv \log\left(\frac{a_{end}}{a}\right)=\int_{t}^{t_{end}}H\mathrm{d}t \approx \int_{\phi_{end}}^{\phi} \frac{1}{\sqrt{2\epsilon_V}}\frac{\mathrm{d}\phi}{M_{Pl}}.
\end{equation} 
The Cosmic Microwave Background fluctuations are created about $40$ to $60$ e-folds before the end of inflation \cite{efolds}. The precise value depends on the details of reheating, post-inflationary thermal history and the energy scale of inflation. The integral constraint $N(\phi_{cmb})\approx 40 - 60$ provides us with the field value when the CMB fluctuations are created. This in turn can be used to find the main observables related to the power spectrum of the fluctuations on the sky:
\begin{equation}\label{eq:observ}
\begin{aligned}
A_{s}&\approx \frac{V}{24\pi^2 M_{Pl}^4 \epsilon_{V}},\\
n_{s} -1 &\approx 2\eta_V-6\epsilon_V,\\
r &\approx 16\epsilon_V,
\end{aligned}
\end{equation}
where the scalar amplitude $A_s$, the scalar spectral index $n_s$, and the tensor-to-scalar ratio $r$ are evaluated at $\phi_{cmb}$. 

\begin{figure}[t]
\includegraphics[width=\columnwidth]{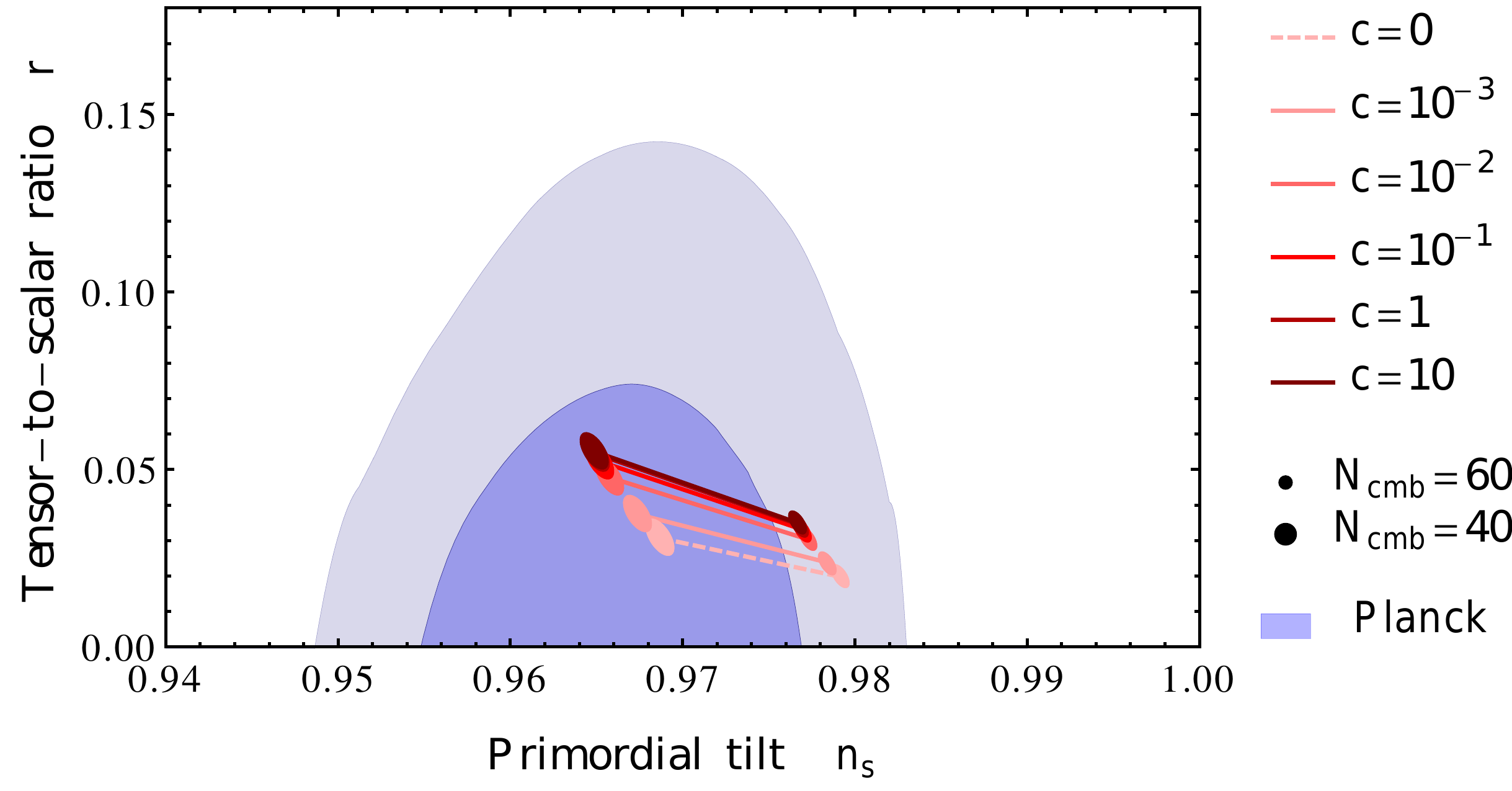}
\caption{Predictions for the $(n_s,r)$ plane for $-10^{-1} \lesssim E < 0$ against the observations (\textit{Planck} TT+lowP data).}
\label{fig:rvsnseps}
\end{figure}

We see from Eqs. \eqref{eq:slowroll} and \eqref{eq:observ} that $n_s$ and $r$ do not depend on the overall factor $M^4$. Scanning the interesting parameter ranges, $-10^{-1}\lesssim E<0$ and $0\leq c<\infty $, the primordial tilt $n_s$ changes by at most $0.5$\% and the tensor-to-scalar ratio $r$ by a couple of per cent. Thus the slow-roll results stemming from the potentials considered here are fairly universal as well as completely consistent with the current observational constraints coming from \textit{Planck} \citep{planck2015}. We plot the predictions for the $(n_s,r)$ pair in Fig. \ref{fig:rvsnseps}.

On the plateau, the potential of the field is $V\sim |E| M^4$. As a result the amplitude of the scalar perturbations $A_s \sim M^4 |E| / M_{Pl}^4$ is degenerate in parameters $E$ and $M$. Observationally, the amplitude is given by $A_s\approx2.3\times10^{-9} $. For the ranges of $E$ and $c$ considered here, $M$ can range from about two orders of magnitude below the GUT scale all the way to the Planck scale where our formalism breaks down as there is very little RG flow and gravitational effects become important. As a limiting case the potential in Eq. \eqref{eq:epspotential} gives the correct size of $A_s$ for the combination $M \sim M_{Pl}$ and $E \sim -10^{-10}$.

\subsection{The discrete case with $E=-1$}
In the remainder of this paper, we will limit ourselves to the theory with $E=-1$ in Eq. \eqref{eigen1}. As argued above, this special choice of the eigenvalue corresponds to the IR theory that is connected to the free theory. In terms of our quantum mechanical problem, this is the un-normalisable solution that contains the lowest-state normalisable mode with a discrete spectrum of integral eigenvalues. Eq. \eqref{SolPsi} now gives us the potential for $\phi$, which is
\begin{align}
\frac{V(\phi)}{M^4} =&   \, \frac{1}{2} \left(\frac{\phi}{M_{Pl}}\right)^2  \nn
& - \log\left[ 1+ c \sqrt{\frac{\pi}{2}} \, \text{erfi}\left( \frac{\phi}{\sqrt{2} M_{Pl}}\right) \right] + \frac{C(c)}{M^4}. \label{eq:potential} 
\end{align}
We expect the overall scale $M$ to be low, as we are examining the dominant solution of the ERG in the IR. As in the general case, fixing the constant $C$ makes the potential only depend on the overall scale parameter $M$ and the shape parameter $c$.
\begin{figure}
\includegraphics[width=\columnwidth]{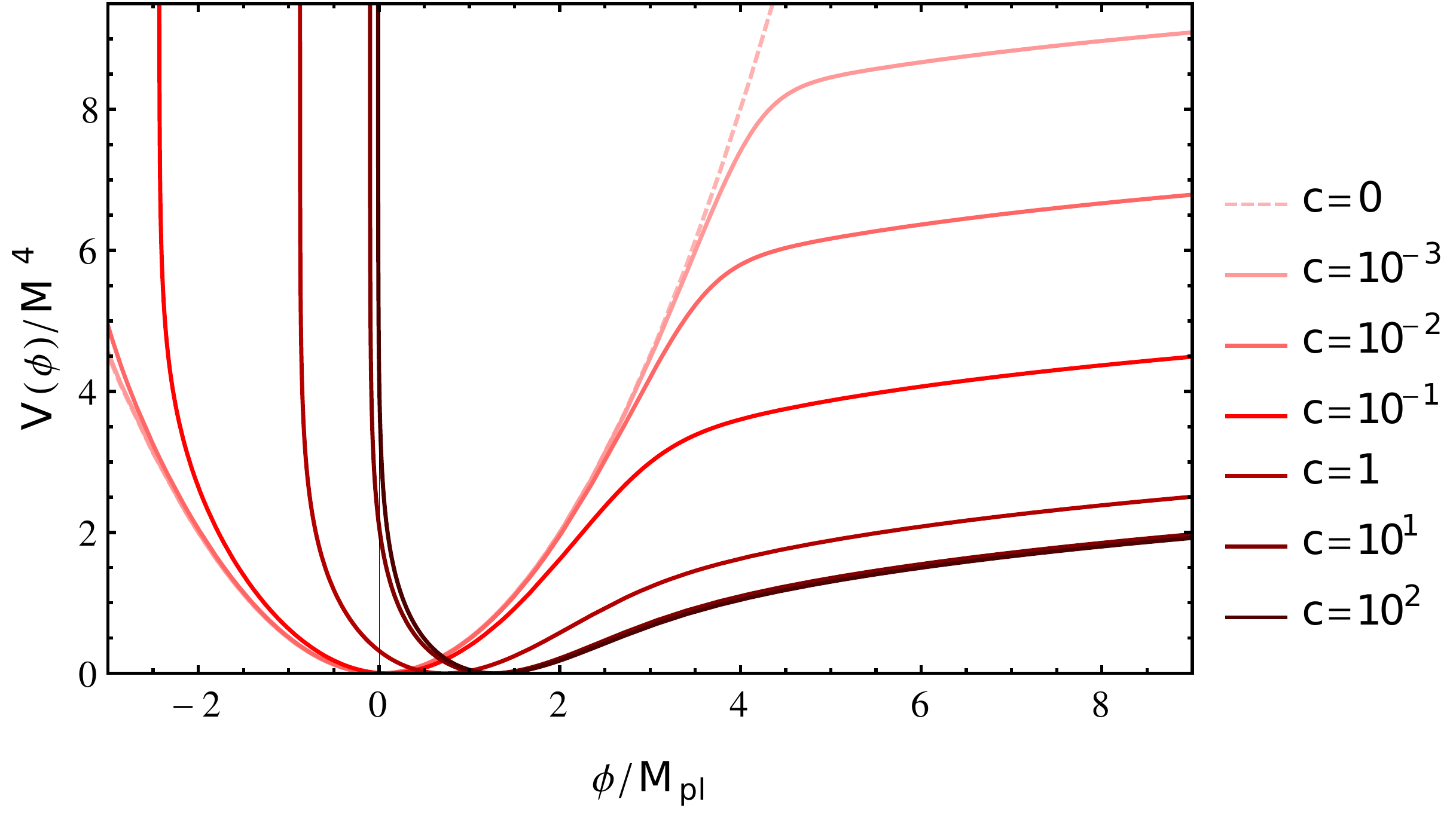}
\caption{The potential from Eq. (\ref{eq:potential}) depending on the parameter $c$. Note that the shape doesn't change for $c \gtrsim \ord{1}$ and that for small $c$ a kink appears in the potential.}
\label{fig:potentials}
\end{figure}

Again, at the minimum of the potential, $\p_\phi V (\phi_{min})=0$, the mass of the scalar field is independent of the shape parameter $c$, 
\begin{align}
\p^2_\phi V (\phi_{min})=M^4/ M_{Pl}^2 .
\end{align} 
Furthermore, when $c\gtrsim \ord{1}$, the potential and its derivatives do not depend on $c$ and the potential has a stable shape (see Fig. \ref{fig:potentials}). The plateau of the potential and the quadratic behaviour near the minimum are also fairly independent of $c$. The parameter governs how sharp the transition is between the two regimes and where it happens. For smaller $c$ the transition happens at higher $\phi$. At $c=0$ we restore the familiar quadratic potential of the free theory.

In the plateau region we use the asymptotic expansion of the error function, 
\begin{align}
\sqrt{\frac{\pi}{2}} \, \text{erfi} \left(\frac{\phi}{\sqrt{2}}\right)\sim e^{\phi^2/2} \left(\frac{1}{\phi} + \ldots\right),
\end{align}
which is approximately valid for $\phi\gtrsim 4 M_{Pl}$. As before, in Eq. \eqref{eq:logarithmic}, we recover the logarithmic behaviour reminiscent of the radiative corrections to a flat potential. In the special case considered here ($E=-1$), the corrections can be turned off for $c=0$, when the free Gaussian theory is recovered.

\begin{figure}
\includegraphics[width=\columnwidth]{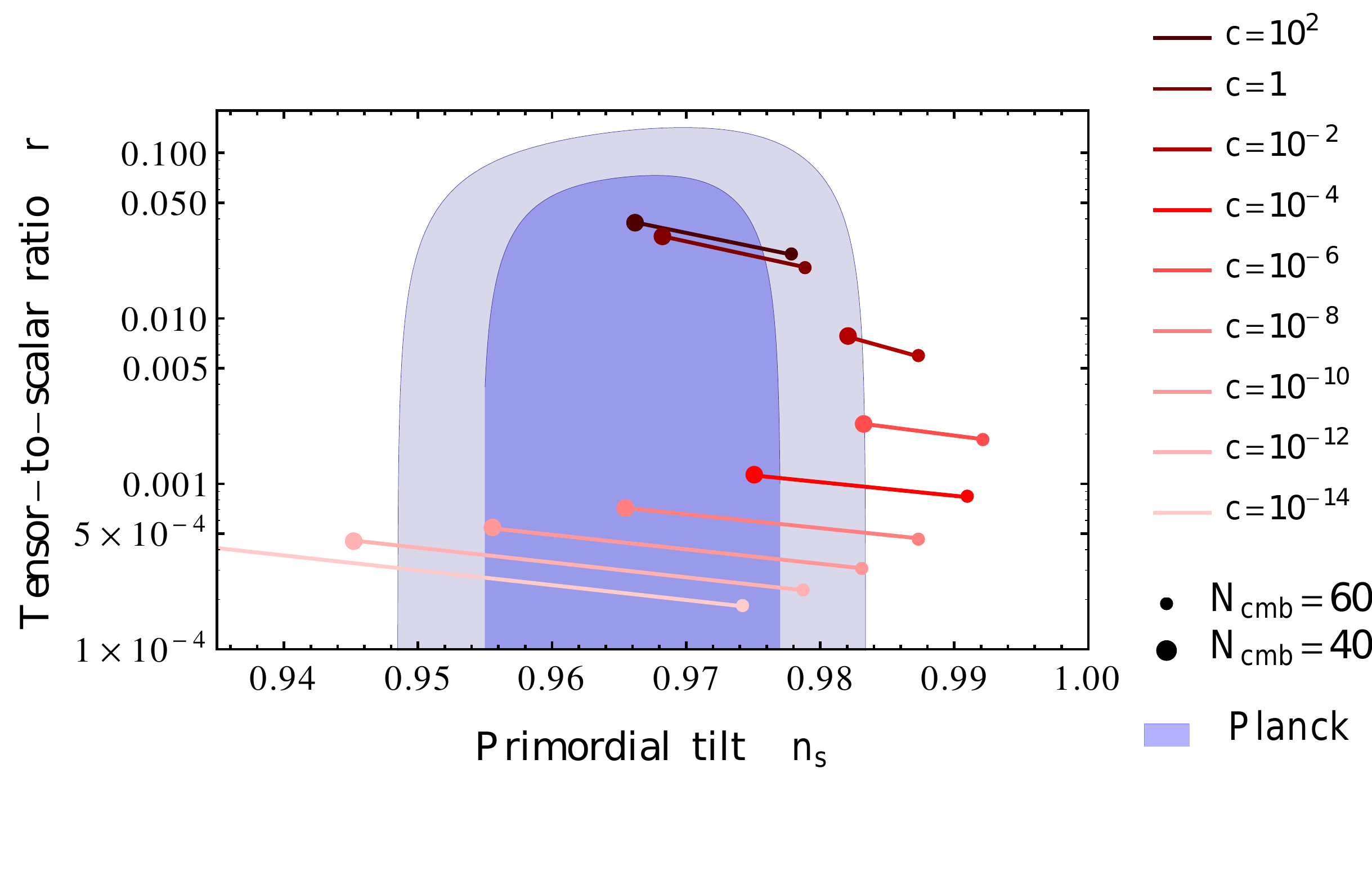}
\caption{Predictions for the $(n_s,r)$ plane against the observations (\textit{Planck} TT+lowP data) when $E=-1$.}
\label{fig:rvsns}
\end{figure}

The slow-roll predictions of our model are plotted in Fig. \ref{fig:rvsns} against the observational constraints. The spectral index and the tensor-to-scalar ratio are independent of the overall scale $M$. For the given shape of the potential (i.e. fixed $c$) the values of $n_s$ and $r$ are determined. Then $M$ is determined such that the amplitude of fluctuations matches the observations. Hence, the inflationary model described by the potential in Eq. (\ref{eq:potential}) would fall in the class of `One parameter' models \citep{encyclinf}.

For the `theoretically preferred' values of the shape parameter, $c\sim\ord{1}$, we are in very good agreement with the observations. Parameter $c$ measures the deformation of the potential away from the free theory ($c=0$). For the small values ($10^{-8}\lesssim c \lesssim 10^{-14}$) the scalar to tensor ratio $r$ can be reduced by several orders of magnitude while being consistent with the observations. This results in the further lowering of the scale of the potential, $M$, to about three orders of magnitude below the GUT scale. This provides more room for the RG flow and adds to the validity of our analysis.

For the cases of small $c$ the potential we study develops a kink (see Fig. \ref{fig:potentials}) that introduces non-negligible derivatives of the field $\phi$. This makes the slow-roll prediction less accurate. The transient violation of the slow-roll conditions in the potential can give rise to observable features in the primordial power spectrum. This motivates further study with the exact integration of the inflationary background and perturbations together with the relaxation of the mean-field approximation in our derivation of the potential.

\section{Conclusion}
In this paper, we studied non-perturbative scalar field potentials derived as an IR limit of the exact renormalisation group equation within the mean-field approximation. We demonstrated that this approximation is valid when considering the potentials of an almost constant field. This setup is precisely what is required to start the exponential expansion of space -- inflation. The potentials we derived were capable of supporting slow-roll inflation with the values of the spectral index and tensor-to-scalar ratio fully consistent with the recent observations. Despite our very general treatment of the scalar field potentials, the slow-roll results are largely independent on the constants parametrising the potentials and thus result in fairly universal predictions. For the special case where we required that the IR theory was continuously connected to a free Gaussian fixed point, we identified a parameter range that led to transient violation of the slow-roll conditions. Hence, this results in a possibly observable feature in the primordial power spectrum, which should be further studied in the future.

\section*{Acknowledgements}
The authors would like to thank Atish Dabholkar, Edward Hardy, Mark Hertzberg, Janos Polonyi, Subir Sarkar and Koenraad Schalm for useful conversations and correspondence. SG is supported in part by a VICI grant of the Netherlands Organization for Scientific Research (NWO), and by the Netherlands Organization for Scientific Research/Ministry of Science and Education (NWO/OCW). DK is supported by the STFC. EES is supported by the ILP LABEX (under reference ANR-10-LABX-63), and by French state funds managed by the ANR within the Investissements dAvenir program under reference ANR-11-IDEX-0004-02.

\begin{appendix}
\section{Decomposition of the UV theory in the mean-field approximation for scalar theories}\label{app:completeset}
In this appendix, we consider in greater detail the decomposition of UV scalar theories in terms of the ERG equation {\em basis} presented in Eq. \eqref{CompleteSetExp}. We will assume the space-time derivatives of $\phi$ to be very small, so that we can use the mean-field approximation in which we can treat $\phi$ as a constant real variable $x$. In this approximation, the path integral \eqref{UVPathIntegral} becomes a one-dimensional integral
\begin{align}\label{ZMFA}
Z = \int \CD \phi \, \Psi_{UV} \left[\phi\right] \approx \int_{-\infty}^{\infty} dx \, e^{ - S(x) },
\end{align}
where the Euclidean action $S(x)$ equals the potential $V(x)$ of the UV scalar theory. 

In order for the expression \eqref{ZMFA} to be well-defined, we assume that the theory is normalisable, i.e. that the integral over $x$ in \eqref{ZMFA} is finite, cf. Eq. \eqref{DefNormal}. It is important to note that in a general quantum field theory, i.e. in the absence of a precise definition of the path integral, this need not be the case. All that is usually assumed is that the potential is bounded from below. As a simple example of a normalisable theory, we can think of the $\phi^4$ theory, which gives
\begin{align}\label{phi4Z}
Z_{\phi^4} &= \int_{-\infty}^{\infty} dx\, e^{-\frac{1}{2} m^2 x^2 - \frac{1}{4!} \lambda x^4 } \nn
&= \sqrt{ \frac{3m^2}{\lambda}  } e^{\frac{3m^4}{4\lambda}} K_{1/4} \left( \frac{3m^4}{4\lambda}\right).
\end{align}
In the expression, $K_\alpha(x)$ is the modified Bessel function of the second kind. In the small $\lambda$ and $m$ expansions,
\begin{align}
&Z_{\phi^4} = \frac{\sqrt{2\pi}}{m} - \sqrt{\frac{\pi}{2}} \frac{\lambda}{4m^5} + \CO\left(\lambda^2 / m^9\right),\\ 
&Z_{\phi^4} = \Gamma\left(\frac{1}{4}\right) \left( \frac{3}{2\lambda} \right)^{1/4}  +\CO\left(m^2 / \lambda^{3/4} \right),
\end{align}
respectively. The expansions imply that at $m=0$, a perturbative treatment around $\lambda = 0$ would be ill-defined as the partition function is divergent.

As an example of a well-known and well-defined QFT, which is ``un-normalisable" from the point of view of our analysis, one can think of the Sine-Gordon potential. In this case,  
\begin{align}
Z_{SG} = \int_{-\infty}^{\infty} dx\, \exp\left\{\frac{m^4}{\lambda} \left[ \cos\left( \frac{\sqrt{\lambda} x}{m}  \right) - 1\right]  \right\} 
\end{align}
is divergent and a more general approach would be needed to treat such theories in the MFA of the ERG. In such cases, the simple procedure of Eq. \eqref{ZMFA} is insufficient for computational purposes. 

To show how the computation of the ERG-basis decomposition \eqref{CompleteSetExp} can be performed in simple example of scalar UV theories, we restrict our attention to normalisable theories with polynomial potentials and consider a single scalar field theory in the MFA with
\begin{align}\label{AppAction}
S(x) = \frac{1}{2} m^2 x^2 + \CU(x) .
\end{align} 
We assume that $S(x)$ gives a finite $Z$, as for example in the $\phi^4$ theory case computed in \eqref{phi4Z}, for which $\CU(x) =  \lambda x^4 / 4!$.

The general solution to the ERG equation \eqref{MeanRG2} was presented in Eq. \eqref{SolPsiGeneral}. For completeness, we restate it here:
\begin{align}\label{SolPsiGeneralApp}
\Psi (x) =&\, e^{-\frac{1}{2}x^2} \biggr[ C_0\, {}_{1}F_{1} \left( \frac{1+E}{2}; \frac{1}{2}; \frac{x^2}{2} \right) \nn
&+ D_0 \,x \sqrt{\frac{2}{\pi}}  \,{}_{1}F_{1}\left( \frac{2+E}{2}; \frac{3}{2}; \frac{x^2}{2} \right) \biggr].
\end{align}
To form a complete set of functions out of \eqref{SolPsiGeneralApp}, we will restrict ourselves to the basis of normalisable solutions for which $E$ are negative integers, i.e. $E \in \{-1, -2, -3, \ldots \}$, with $C_0 = 0$ for $E\in\{-2,-4,\ldots\}$ and $D_0 = 0 $ for $E\in\{-1,-3,\ldots\}$. Since the first parameter of the hypergeometric function is now always an integer, it is useful to rewrite the confluent hypergeometric functions in terms of the Hermite polynomials:
\begin{align}
&{}_{1}F_{1} \left( - n ; \frac{1}{2} ; \frac{x^2}{2} \right) = \frac{(-1)^n n!}{(2n)!} H_{2n} \left(\frac{x}{\sqrt{2} }\right) , \\
&\sqrt{2} \,x\, {}_{1}F_{1} \left( - n ; \frac{3}{2} ; \frac{x^2}{2} \right) = \frac{(-1)^n n!}{(2n+1)!} H_{2n+1} \left(\frac{x}{\sqrt{2} }\right).
\end{align}
The Hermite polynomials multiplied by $\exp \{-x^2/4\}$, i.e. the functions in the set of
\begin{align}\label{HermiteBasis}
\left\{ e^{-x^2 / 4 } H_n \left(\frac{x}{\sqrt{2}} \right) \right\},~~~n \geq 0,
\end{align}
form a complete orthonormal basis on the Hilbert space of $L^2$-integrable functions on the entire real axis $x \in (-\infty,\infty)$. The completeness relation has the form\footnote{For a detailed discussion of Hermite and Laguerre polynomials, their orthogonality relation and the proof of completeness, see Chapter V. of the reference \cite{szego1939orthogonal}.}
\begin{align}\label{LaguerreComp}
\int_{-\infty}^\infty dx \, e^{-x^2/2} H_n \left( \frac{x}{\sqrt{2}}\right) H_m \left( \frac{x}{\sqrt{2}}\right) = \sqrt{2 \pi} 2^n n! \, \delta_{nm},
\end{align}
where $n,m \in \{0,1,2,\ldots\}$. Using the Hermite polynomial basis, we can now write the ERG flows of normalisable $\Psi_{UV}$ theories as 
\begin{align}\label{PsiExp}
\Psi_{UV} (t,x) = \sum_{n=0}^\infty \gamma_n e^{- t (n+1)} e^{-\frac{1}{2} x^2} H_n \left( \frac{x}{\sqrt{2}} \right) .
\end{align}
Note also that $\Psi_{UV} (x) = \Psi_{UV} (t=0,x)$. Using the completeness relation \eqref{LaguerreComp}, we can express the coefficients $\gamma_n$ of the $\Psi_{UV}$ expansion (computed at $t=0$) as
\begin{align}
\gamma_n = \frac{1}{\sqrt{2\pi} 2^n n!} \int_{-\infty}^\infty dx \,\Psi_{UV} (x) H_n \left( \frac{x}{\sqrt{2}} \right) .
\end{align}
In our prototypical example of the $\phi^4$ theory, for which $\Psi_{UV}$ is even in $x$ (and $y$), it is clear that only $\gamma_n$ with even $n$ contribute to $\Psi_{UV}$. Similarly, this would be true in any theory with an even $\CU(x)$. To solve for the $\gamma_n$ coefficients in such theories, it is particularly convenient to write the Hermite polynomials in terms of associated Laguerre polynomials,
\begin{align}
&H_{2n} (y) = (-1)^n 2^{2n} n! \, L_n^{(-1/2)} \left(y^2\right), \\
&H_{2n+1} (y) = (-1)^n 2^{2n+1} n! \, L_n^{(1/2)} \left(y^2\right),
\end{align}
since they obey the recurrence relation
\begin{align}\label{Recurr}
L^{(\alpha)}_{n+1}\left(y^2 \right) =&  \left( \frac{ 2n+1 + \alpha - y^2 }{n+1} \right) L^{(\alpha)}_{n}  \left(y^2 \right) \nn
&- \left( \frac{ n+\alpha  }{n+1} \right) L^{(\alpha)}_{n-1}\left(y^2\right) .
\end{align}
Note that $y = x / \sqrt{2}$. Using Eq. \eqref{Recurr} with $\alpha=-1/2$ and $\alpha=1/2$, we can find the recurrence relations for {\em all} even and odd $\gamma_n$ from the knowledge of $\left( \gamma_0, \gamma_2\right)$ and $\left(\gamma_1,\gamma_3\right)$, respectively:
\begin{align}
\gamma_{n+2} =&~ \frac{1}{(n+2)(n+1)} \left[ \frac{\partial \gamma_n }{\partial m^2} \right.\nn
&\left. + \left(n+\frac{1}{2}\right) \gamma_n  + \frac{1}{4} \gamma_{n-2} \right] .
\end{align}

To see how this works on a specific example, let us again return to the case of the $\phi^4$ theory, for which
\begin{align}\label{GammaRec}
\gamma_0 =&~ \sqrt{\frac{3m^2}{2\pi\lambda}} e^{\frac{3m^4}{4\lambda}} K_{1/4} \left( \frac{3m^4}{4\lambda}  \right) ,\\
\gamma_2 =&~ \frac{1}{8} \sqrt{\frac{3\pi}{m^2\lambda^3}} e^{\frac{3m^4}{4\lambda} }   \left[ \left(  3m^4 + \left(2+m^2\right)\lambda \right) I_{1/4}\left( \frac{3m^4}{4\lambda}  \right)   \right. \nn
&\left. - m^2 \left(3m^2+\lambda\right) I_{-1/4}\left( \frac{3m^4}{4\lambda}  \right) \right. \nn
&\left. +3 m^4 \left(I_{5/4}\left( \frac{3m^4}{4\lambda}  \right) - I_{3/4}\left( \frac{3m^4}{4\lambda}  \right) \right) \right]. 
\end{align}
The rest of the coefficients $\gamma_n$ can then be generated using \eqref{GammaRec}. Note that $I_\alpha(x)$ is the modified Bessel function of the first kind. As a check on the results, we note that in the UV (at $t=0$), we need to integrate over the entire real axis, $x\in(-\infty,\infty)$. In that case, only the $n=0$ term contributes to the integral over the sum \eqref{PsiExp}. In the extreme IR limit of $t\to\infty$, the dominant contribution to the RG flow of $\Psi_{UV}(t,x)$ comes from the $n=0$ (or $E=-1$) term, i.e. the Gaussian effective action.

We note that the effective IR potential of the $\phi^4$ theory is not of the form we used to study inflation. This is because in this appendix we restricted ourselves to only the basis of hypergeometric functions for which the first parameter was an integer. In case of the $\phi^4$ theory and other normalisable theories, this is sufficient to decompose the entire theory and solve the RG flow. However, we expect that for more general theory, in particular those with un-normalisable potentials, this type of a decomposition would not suffice and a more general basis would be required.

\section{Momentum Space Solution}
\label{app:Momspace}
In this appendix, we solve the system in momentum space, where the solution takes a much simpler form. We first recall the equation 
\begin{equation}
\label{eq:rgeqApp}
\p_{\tilde t}\Psi(x)=\hat H\Psi(x)=\left(\p_x^2+x\p_x\right)\Psi(x),
\end{equation}
where $\tilde t$ is the appropriately rescaled RG time. Note that the Hamiltonian is symmetric in $x$, so that solutions decompose into symmetric solutions $\Psi_s(x)$ and anti-symmetric solutions $\Psi_a(x)$, respectively.

Let us go to momentum space and write
\begin{equation}
\Psi(x)=\int_p\Psi(p)e^{ipx},
\end{equation}
where
\begin{equation}
\int_p=\frac{1}{2\pi}\int_{-\infty}^\infty\d p.
\end{equation}
Plugging this into Eq. \eqref{eq:rgeqApp}, we find
\begin{equation}
\int_p\left(\p_{\tilde t}\Psi(p)+p^2\Psi(p)-\Psi(p)p\,\p_p\right)e^{ipx}=0.
\end{equation}
This gives the RG equation in momentum space after an integration by parts on the last term,
\begin{equation}
\p_{\tilde t}\Psi(p)=-\left(p^2+1+p\,\p_p\right)\Psi(p).
\end{equation}
Note that the equation is a first-order differential equation. Let us perform a redefinition of coordinates,
\begin{equation}
p\,\p_p=\p_q,
\end{equation}
which is satisfied by
\begin{equation}
p=K\,e^q\:,
\end{equation}
for some constant $K$. We can set $K=1$ without loss of generality. The RG equation then reads
\begin{equation}
\p_{\tilde t}\Psi(q)=-\left(e^{2q}+1+\p_q\right)\Psi(q).
\end{equation}
Rescaling the theory as
\begin{equation*}
\tilde\Psi(q)= \exp \left(\frac{1}{2}\,e^{2q}+q\right)\Psi(q),
\end{equation*}
we find
\begin{equation}
\label{eq:RGeqReScale}
\p_{\tilde t}\tilde\Psi(q)=-\p_q\tilde\Psi(q).
\end{equation}
We can now expand $\tilde\Psi(q)$ as usual in a basis $\{\psi_k=e^{-E_kq}\}$,
\begin{equation}
\label{eq:expE}
\tilde\Psi(q)=\int_E\,C(E)\,e^{-Eq}\:.
\end{equation}
Note that $\tilde\Psi(q)$ is closely related to the Laplace transform of $C(E)$. Indeed, we will see below that we need $C(E)=0$ for $E\ge0$. The basis $\{\psi_k=e^{-E_kq}\}$ is orthogonal with respect to the inner product
\begin{equation}
(\psi_1,\psi_2)=\frac{1}{2\pi i}\int_{-i\infty}^{i\infty}\d q \,e^{q(E_1-E_2)}=\delta(E_1-E_2).
\end{equation}
Note that in terms of $p$, restricting ourselves to $E\in\mathbb{Z}$, the expansion \eqref{eq:expE} is just the regular Laurent expansion of $\tilde\Psi(p)$.

We can now solve the re-defined RG equation \eqref{eq:RGeqReScale}. That is, we assume that the UV theory has the expansion
\begin{equation}
\tilde\Psi_{UV}(p)=\int_EC(E)p^{-E}.
\end{equation}
The solution of \eqref{eq:RGeqReScale} then reads
\begin{equation}
\tilde\Psi_t(p)=\int_EC(E)\,e^{tE}p^{-E}.
\end{equation}
The ``eigen-theories" are therefore given by
\begin{equation}
\label{eq:MomEgen}
\Psi(p)=\frac{e^{-\frac{1}{2}p^2}}{p^{1+E}}.
\end{equation}
As $t\rightarrow\infty$, we expect that the eigen-theories with largest $E$, for which $C(E)\neq0$, will dominate. Fourier transforming \eqref{eq:MomEgen} back to position space and taking the symmetric part then gives
\begin{equation}
\Psi_s(x)\propto e^{-\frac{1}{2}x^2}{}_{1}F_{1} \left( \frac{1+E}{2}; \frac{1}{2}; \frac{x^2}{2} \right),
\end{equation}
while taking the anti-symmetric part gives
\begin{equation}
\Psi_a(x)\propto x\,e^{-\frac{1}{2}x^2}{}_{1}F_{1}\left( \frac{2+E}{2}; \frac{3}{2}; \frac{x^2}{2} \right),
\end{equation}
as expected. Moreover, we find that the inverse transform is ill-defined for $E\ge0$. That is, we require that the UV theory has $C(E)=0$ for $E\ge0$.

\end{appendix}

\bibliography{Universal_inflationBIB}

\begin{thebibliography}{36}%
\makeatletter
\providecommand \@ifxundefined [1]{%
 \@ifx{#1\undefined}
}%
\providecommand \@ifnum [1]{%
 \ifnum #1\expandafter \@firstoftwo
 \else \expandafter \@secondoftwo
 \fi
}%
\providecommand \@ifx [1]{%
 \ifx #1\expandafter \@firstoftwo
 \else \expandafter \@secondoftwo
 \fi
}%
\providecommand \natexlab [1]{#1}%
\providecommand \enquote  [1]{``#1''}%
\providecommand \bibnamefont  [1]{#1}%
\providecommand \bibfnamefont [1]{#1}%
\providecommand \citenamefont [1]{#1}%
\providecommand \href@noop [0]{\@secondoftwo}%
\providecommand \href [0]{\begingroup \@sanitize@url \@href}%
\providecommand \@href[1]{\@@startlink{#1}\@@href}%
\providecommand \@@href[1]{\endgroup#1\@@endlink}%
\providecommand \@sanitize@url [0]{\catcode `\\12\catcode `\$12\catcode
  `\&12\catcode `\#12\catcode `\^12\catcode `\_12\catcode `\%12\relax}%
\providecommand \@@startlink[1]{}%
\providecommand \@@endlink[0]{}%
\providecommand \url  [0]{\begingroup\@sanitize@url \@url }%
\providecommand \@url [1]{\endgroup\@href {#1}{\urlprefix }}%
\providecommand \urlprefix  [0]{URL }%
\providecommand \Eprint [0]{\href }%
\providecommand \doibase [0]{http://dx.doi.org/}%
\providecommand \selectlanguage [0]{\@gobble}%
\providecommand \bibinfo  [0]{\@secondoftwo}%
\providecommand \bibfield  [0]{\@secondoftwo}%
\providecommand \translation [1]{[#1]}%
\providecommand \BibitemOpen [0]{}%
\providecommand \bibitemStop [0]{}%
\providecommand \bibitemNoStop [0]{.\EOS\space}%
\providecommand \EOS [0]{\spacefactor3000\relax}%
\providecommand \BibitemShut  [1]{\csname bibitem#1\endcsname}%
\let\auto@bib@innerbib\@empty
\bibitem [{\citenamefont {Guth}(1981)}]{PhysRevD.23.347}%
  \BibitemOpen
  \bibfield  {author} {\bibinfo {author} {\bibfnamefont {A.~H.}\ \bibnamefont
  {Guth}},\ }\href {\doibase 10.1103/PhysRevD.23.347} {\bibfield  {journal}
  {\bibinfo  {journal} {Phys. Rev. D}\ }\textbf {\bibinfo {volume} {23}},\
  \bibinfo {pages} {347} (\bibinfo {year} {1981})}\BibitemShut {NoStop}%
\bibitem [{\citenamefont {{Starobinsky}}(1980)}]{starinf}%
  \BibitemOpen
  \bibfield  {author} {\bibinfo {author} {\bibfnamefont {A.~A.}\ \bibnamefont
  {{Starobinsky}}},\ }\href {\doibase 10.1016/0370-2693(80)90670-X} {\bibfield
  {journal} {\bibinfo  {journal} {Physics Letters B}\ }\textbf {\bibinfo
  {volume} {91}},\ \bibinfo {pages} {99} (\bibinfo {year} {1980})}\BibitemShut
  {NoStop}%
\bibitem [{\citenamefont {Linde}(1982)}]{Linde:1981mu}%
  \BibitemOpen
  \bibfield  {author} {\bibinfo {author} {\bibfnamefont {A.~D.}\ \bibnamefont
  {Linde}},\ }\bibfield  {booktitle} {\emph {\bibinfo {booktitle} {{In *Moscow
  1981, Proceedings, Quantum Gravity*, 185-195 and Moscow Inst. Phys. Acad.
  Sci. - 81-229 (81,REC.DEC.) 15p}}},\ }\href {\doibase
  10.1016/0370-2693(82)91219-9} {\bibfield  {journal} {\bibinfo  {journal}
  {Phys. Lett.}\ }\textbf {\bibinfo {volume} {B108}},\ \bibinfo {pages} {389}
  (\bibinfo {year} {1982})}\BibitemShut {NoStop}%
\bibitem [{\citenamefont {{Mukhanov}}\ and\ \citenamefont
  {{Chibisov}}(1981)}]{mukhfluct}%
  \BibitemOpen
  \bibfield  {author} {\bibinfo {author} {\bibfnamefont {V.~F.}\ \bibnamefont
  {{Mukhanov}}}\ and\ \bibinfo {author} {\bibfnamefont {G.~V.}\ \bibnamefont
  {{Chibisov}}},\ }\href@noop {} {\bibfield  {journal} {\bibinfo  {journal}
  {Soviet Journal of Experimental and Theoretical Physics Letters}\ }\textbf
  {\bibinfo {volume} {33}},\ \bibinfo {pages} {532} (\bibinfo {year}
  {1981})}\BibitemShut {NoStop}%
\bibitem [{\citenamefont {{Lyth}}\ and\ \citenamefont
  {{Liddle}}(2009)}]{inflbook}%
  \BibitemOpen
  \bibfield  {author} {\bibinfo {author} {\bibfnamefont {D.~H.}\ \bibnamefont
  {{Lyth}}}\ and\ \bibinfo {author} {\bibfnamefont {A.~R.}\ \bibnamefont
  {{Liddle}}},\ }\href@noop {} {\emph {\bibinfo {title} {The Primordial Density
  Perturbation, by David H.~Lyth , Andrew R.~Liddle, Cambridge, UK: Cambridge
  University Press, 2009}}}\ (\bibinfo {year} {2009})\BibitemShut {NoStop}%
\bibitem [{\citenamefont {{Martin}}\ \emph {et~al.}(2014)\citenamefont
  {{Martin}}, \citenamefont {{Ringeval}},\ and\ \citenamefont
  {{Vennin}}}]{encyclinf}%
  \BibitemOpen
  \bibfield  {author} {\bibinfo {author} {\bibfnamefont {J.}~\bibnamefont
  {{Martin}}}, \bibinfo {author} {\bibfnamefont {C.}~\bibnamefont
  {{Ringeval}}}, \ and\ \bibinfo {author} {\bibfnamefont {V.}~\bibnamefont
  {{Vennin}}},\ }\href {\doibase 10.1016/j.dark.2014.01.003} {\bibfield
  {journal} {\bibinfo  {journal} {Physics of the Dark Universe}\ }\textbf
  {\bibinfo {volume} {5}},\ \bibinfo {pages} {75} (\bibinfo {year} {2014})},\
  \Eprint {http://arxiv.org/abs/1303.3787} {arXiv:1303.3787 [astro-ph.CO]}
  \BibitemShut {NoStop}%
\bibitem [{\citenamefont {Salvio}\ and\ \citenamefont
  {Strumia}(2014)}]{Salvio:2014soa}%
  \BibitemOpen
  \bibfield  {author} {\bibinfo {author} {\bibfnamefont {A.}~\bibnamefont
  {Salvio}}\ and\ \bibinfo {author} {\bibfnamefont {A.}~\bibnamefont
  {Strumia}},\ }\href {\doibase 10.1007/JHEP06(2014)080} {\bibfield  {journal}
  {\bibinfo  {journal} {JHEP}\ }\textbf {\bibinfo {volume} {06}},\ \bibinfo
  {pages} {080} (\bibinfo {year} {2014})},\ \Eprint
  {http://arxiv.org/abs/1403.4226} {arXiv:1403.4226 [hep-ph]} \BibitemShut
  {NoStop}%
\bibitem [{\citenamefont {Kannike}\ \emph {et~al.}(2015)\citenamefont
  {Kannike}, \citenamefont {Hütsi}, \citenamefont {Pizza}, \citenamefont
  {Racioppi}, \citenamefont {Raidal}, \citenamefont {Salvio},\ and\
  \citenamefont {Strumia}}]{Kannike:2015apa}%
  \BibitemOpen
  \bibfield  {author} {\bibinfo {author} {\bibfnamefont {K.}~\bibnamefont
  {Kannike}}, \bibinfo {author} {\bibfnamefont {G.}~\bibnamefont {Hütsi}},
  \bibinfo {author} {\bibfnamefont {L.}~\bibnamefont {Pizza}}, \bibinfo
  {author} {\bibfnamefont {A.}~\bibnamefont {Racioppi}}, \bibinfo {author}
  {\bibfnamefont {M.}~\bibnamefont {Raidal}}, \bibinfo {author} {\bibfnamefont
  {A.}~\bibnamefont {Salvio}}, \ and\ \bibinfo {author} {\bibfnamefont
  {A.}~\bibnamefont {Strumia}},\ }\href {\doibase 10.1007/JHEP05(2015)065}
  {\bibfield  {journal} {\bibinfo  {journal} {JHEP}\ }\textbf {\bibinfo
  {volume} {05}},\ \bibinfo {pages} {065} (\bibinfo {year} {2015})},\ \Eprint
  {http://arxiv.org/abs/1502.01334} {arXiv:1502.01334 [astro-ph.CO]}
  \BibitemShut {NoStop}%
\bibitem [{\citenamefont {Copeland}\ \emph {et~al.}(2015)\citenamefont
  {Copeland}, \citenamefont {Rahmede},\ and\ \citenamefont
  {Saltas}}]{Copeland:2013vva}%
  \BibitemOpen
  \bibfield  {author} {\bibinfo {author} {\bibfnamefont {E.~J.}\ \bibnamefont
  {Copeland}}, \bibinfo {author} {\bibfnamefont {C.}~\bibnamefont {Rahmede}}, \
  and\ \bibinfo {author} {\bibfnamefont {I.~D.}\ \bibnamefont {Saltas}},\
  }\href {\doibase 10.1103/PhysRevD.91.103530} {\bibfield  {journal} {\bibinfo
  {journal} {Phys. Rev.}\ }\textbf {\bibinfo {volume} {D91}},\ \bibinfo {pages}
  {103530} (\bibinfo {year} {2015})},\ \Eprint {http://arxiv.org/abs/1311.0881}
  {arXiv:1311.0881 [gr-qc]} \BibitemShut {NoStop}%
\bibitem [{\citenamefont {Kaya}(2013)}]{Kaya:2013bga}%
  \BibitemOpen
  \bibfield  {author} {\bibinfo {author} {\bibfnamefont {A.}~\bibnamefont
  {Kaya}},\ }\href {\doibase 10.1103/PhysRevD.87.123501} {\bibfield  {journal}
  {\bibinfo  {journal} {Phys. Rev.}\ }\textbf {\bibinfo {volume} {D87}},\
  \bibinfo {pages} {123501} (\bibinfo {year} {2013})},\ \Eprint
  {http://arxiv.org/abs/1303.5459} {arXiv:1303.5459 [hep-th]} \BibitemShut
  {NoStop}%
\bibitem [{\citenamefont {Saltas}(2016)}]{Saltas:2015vsc}%
  \BibitemOpen
  \bibfield  {author} {\bibinfo {author} {\bibfnamefont {I.~D.}\ \bibnamefont
  {Saltas}},\ }\href {\doibase 10.1088/1475-7516/2016/02/048} {\bibfield
  {journal} {\bibinfo  {journal} {JCAP}\ }\textbf {\bibinfo {volume} {1602}},\
  \bibinfo {pages} {048} (\bibinfo {year} {2016})},\ \Eprint
  {http://arxiv.org/abs/1512.06134} {arXiv:1512.06134 [hep-th]} \BibitemShut
  {NoStop}%
\bibitem [{\citenamefont {Serreau}(2014)}]{serreau2014renormalization}%
  \BibitemOpen
  \bibfield  {author} {\bibinfo {author} {\bibfnamefont {J.}~\bibnamefont
  {Serreau}},\ }\href@noop {} {\bibfield  {journal} {\bibinfo  {journal}
  {Physics Letters B}\ }\textbf {\bibinfo {volume} {730}},\ \bibinfo {pages}
  {271} (\bibinfo {year} {2014})}\BibitemShut {NoStop}%
\bibitem [{\citenamefont {Benedetti}(2015)}]{benedetti2015critical}%
  \BibitemOpen
  \bibfield  {author} {\bibinfo {author} {\bibfnamefont {D.}~\bibnamefont
  {Benedetti}},\ }\href@noop {} {\bibfield  {journal} {\bibinfo  {journal}
  {Journal of Statistical Mechanics: Theory and Experiment}\ }\textbf {\bibinfo
  {volume} {2015}},\ \bibinfo {pages} {P01002} (\bibinfo {year}
  {2015})}\BibitemShut {NoStop}%
\bibitem [{\citenamefont {Guilleux}\ and\ \citenamefont
  {Serreau}(2015)}]{Guilleux:2015pma}%
  \BibitemOpen
  \bibfield  {author} {\bibinfo {author} {\bibfnamefont {M.}~\bibnamefont
  {Guilleux}}\ and\ \bibinfo {author} {\bibfnamefont {J.}~\bibnamefont
  {Serreau}},\ }\href@noop {} {\  (\bibinfo {year} {2015})},\ \Eprint
  {http://arxiv.org/abs/1506.06183} {arXiv:1506.06183 [hep-th]} \BibitemShut
  {NoStop}%
\bibitem [{\citenamefont {{Vachaspati}}\ and\ \citenamefont
  {{Trodden}}(2000)}]{causalVachas}%
  \BibitemOpen
  \bibfield  {author} {\bibinfo {author} {\bibfnamefont {T.}~\bibnamefont
  {{Vachaspati}}}\ and\ \bibinfo {author} {\bibfnamefont {M.}~\bibnamefont
  {{Trodden}}},\ }\href {\doibase 10.1103/PhysRevD.61.023502} {\bibfield
  {journal} {\bibinfo  {journal} {Phys. Rev. D}\ }\textbf {\bibinfo {volume}
  {61}},\ \bibinfo {eid} {023502} (\bibinfo {year} {2000})},\ \Eprint
  {http://arxiv.org/abs/gr-qc/9811037} {gr-qc/9811037} \BibitemShut {NoStop}%
\bibitem [{\citenamefont {Starobinsky}(1986)}]{starobinsky1986stochastic}%
  \BibitemOpen
  \bibfield  {author} {\bibinfo {author} {\bibfnamefont {A.~A.}\ \bibnamefont
  {Starobinsky}},\ }in\ \href@noop {} {\emph {\bibinfo {booktitle} {Field
  theory, quantum gravity and strings}}}\ (\bibinfo  {publisher} {Springer},\
  \bibinfo {year} {1986})\ pp.\ \bibinfo {pages} {107--126}\BibitemShut
  {NoStop}%
\bibitem [{\citenamefont {Starobinsky}\ and\ \citenamefont
  {Yokoyama}(1994)}]{Starobinsky:1994bd}%
  \BibitemOpen
  \bibfield  {author} {\bibinfo {author} {\bibfnamefont {A.~A.}\ \bibnamefont
  {Starobinsky}}\ and\ \bibinfo {author} {\bibfnamefont {J.}~\bibnamefont
  {Yokoyama}},\ }\href {\doibase 10.1103/PhysRevD.50.6357} {\bibfield
  {journal} {\bibinfo  {journal} {Phys.Rev.}\ }\textbf {\bibinfo {volume}
  {D50}},\ \bibinfo {pages} {6357} (\bibinfo {year} {1994})},\ \Eprint
  {http://arxiv.org/abs/astro-ph/9407016} {arXiv:astro-ph/9407016 [astro-ph]}
  \BibitemShut {NoStop}%
\bibitem [{\citenamefont {Polonyi}(1996)}]{Polonyi:1994pn}%
  \BibitemOpen
  \bibfield  {author} {\bibinfo {author} {\bibfnamefont {J.}~\bibnamefont
  {Polonyi}},\ }\href {\doibase 10.1006/aphy.1996.0133} {\bibfield  {journal}
  {\bibinfo  {journal} {Annals Phys.}\ }\textbf {\bibinfo {volume} {252}},\
  \bibinfo {pages} {300} (\bibinfo {year} {1996})},\ \Eprint
  {http://arxiv.org/abs/hep-th/9409004} {arXiv:hep-th/9409004 [hep-th]}
  \BibitemShut {NoStop}%
\bibitem [{\citenamefont {Litim}\ and\ \citenamefont
  {Vergara}(2004)}]{Litim:2003kf}%
  \BibitemOpen
  \bibfield  {author} {\bibinfo {author} {\bibfnamefont {D.~F.}\ \bibnamefont
  {Litim}}\ and\ \bibinfo {author} {\bibfnamefont {L.}~\bibnamefont
  {Vergara}},\ }\href {\doibase 10.1016/j.physletb.2003.11.047} {\bibfield
  {journal} {\bibinfo  {journal} {Phys. Lett.}\ }\textbf {\bibinfo {volume}
  {B581}},\ \bibinfo {pages} {263} (\bibinfo {year} {2004})},\ \Eprint
  {http://arxiv.org/abs/hep-th/0310101} {arXiv:hep-th/0310101 [hep-th]}
  \BibitemShut {NoStop}%
\bibitem [{\citenamefont {Zappala}(2012)}]{Zappala:2012wh}%
  \BibitemOpen
  \bibfield  {author} {\bibinfo {author} {\bibfnamefont {D.}~\bibnamefont
  {Zappala}},\ }\href {\doibase 10.1103/PhysRevD.86.125003} {\bibfield
  {journal} {\bibinfo  {journal} {Phys. Rev.}\ }\textbf {\bibinfo {volume}
  {D86}},\ \bibinfo {pages} {125003} (\bibinfo {year} {2012})},\ \Eprint
  {http://arxiv.org/abs/1206.2480} {arXiv:1206.2480 [hep-th]} \BibitemShut
  {NoStop}%
\bibitem [{\citenamefont {Fischer}\ and\ \citenamefont
  {Gies}(2004)}]{Fischer:2004uk}%
  \BibitemOpen
  \bibfield  {author} {\bibinfo {author} {\bibfnamefont {C.~S.}\ \bibnamefont
  {Fischer}}\ and\ \bibinfo {author} {\bibfnamefont {H.}~\bibnamefont {Gies}},\
  }\href {\doibase 10.1088/1126-6708/2004/10/048} {\bibfield  {journal}
  {\bibinfo  {journal} {JHEP}\ }\textbf {\bibinfo {volume} {10}},\ \bibinfo
  {pages} {048} (\bibinfo {year} {2004})},\ \Eprint
  {http://arxiv.org/abs/hep-ph/0408089} {arXiv:hep-ph/0408089 [hep-ph]}
  \BibitemShut {NoStop}%
\bibitem [{\citenamefont {Borchardt}\ and\ \citenamefont
  {Knorr}(2015)}]{Borchardt:2015rxa}%
  \BibitemOpen
  \bibfield  {author} {\bibinfo {author} {\bibfnamefont {J.}~\bibnamefont
  {Borchardt}}\ and\ \bibinfo {author} {\bibfnamefont {B.}~\bibnamefont
  {Knorr}},\ }\href {\doibase 10.1103/PhysRevD.93.089904,
  10.1103/PhysRevD.91.105011} {\bibfield  {journal} {\bibinfo  {journal} {Phys.
  Rev.}\ }\textbf {\bibinfo {volume} {D91}},\ \bibinfo {pages} {105011}
  (\bibinfo {year} {2015})},\ \bibinfo {note} {[Erratum: Phys.
  Rev.D93,no.8,089904(2016)]},\ \Eprint {http://arxiv.org/abs/1502.07511}
  {arXiv:1502.07511 [hep-th]} \BibitemShut {NoStop}%
\bibitem [{\citenamefont {Borchardt}\ and\ \citenamefont
  {Knorr}(2016)}]{Borchardt:2016pif}%
  \BibitemOpen
  \bibfield  {author} {\bibinfo {author} {\bibfnamefont {J.}~\bibnamefont
  {Borchardt}}\ and\ \bibinfo {author} {\bibfnamefont {B.}~\bibnamefont
  {Knorr}},\ }\href@noop {} {\  (\bibinfo {year} {2016})},\ \Eprint
  {http://arxiv.org/abs/1603.06726} {arXiv:1603.06726 [hep-th]} \BibitemShut
  {NoStop}%
\bibitem [{\citenamefont {Wilson}\ and\ \citenamefont
  {Kogut}(1974)}]{Wilson:1973jj}%
  \BibitemOpen
  \bibfield  {author} {\bibinfo {author} {\bibfnamefont {K.}~\bibnamefont
  {Wilson}}\ and\ \bibinfo {author} {\bibfnamefont {J.~B.}\ \bibnamefont
  {Kogut}},\ }\href {\doibase 10.1016/0370-1573(74)90023-4} {\bibfield
  {journal} {\bibinfo  {journal} {Phys.Rept.}\ }\textbf {\bibinfo {volume}
  {12}},\ \bibinfo {pages} {75} (\bibinfo {year} {1974})}\BibitemShut {NoStop}%
\bibitem [{\citenamefont {Polonyi}(2003)}]{Polonyi:2001se}%
  \BibitemOpen
  \bibfield  {author} {\bibinfo {author} {\bibfnamefont {J.}~\bibnamefont
  {Polonyi}},\ }\href {\doibase 10.2478/BF02475552} {\bibfield  {journal}
  {\bibinfo  {journal} {Central Eur.J.Phys.}\ }\textbf {\bibinfo {volume}
  {1}},\ \bibinfo {pages} {1} (\bibinfo {year} {2003})},\ \Eprint
  {http://arxiv.org/abs/hep-th/0110026} {arXiv:hep-th/0110026 [hep-th]}
  \BibitemShut {NoStop}%
\bibitem [{\citenamefont {Bervillier}(2013)}]{Bervillier:2013kda}%
  \BibitemOpen
  \bibfield  {author} {\bibinfo {author} {\bibfnamefont {C.}~\bibnamefont
  {Bervillier}},\ }\href {\doibase 10.5488/CMP.16.23003} {\bibfield  {journal}
  {\bibinfo  {journal} {Condens. Matter Phys.}\ }\textbf {\bibinfo {volume}
  {16}},\ \bibinfo {pages} {23003} (\bibinfo {year} {2013})},\ \Eprint
  {http://arxiv.org/abs/1304.4131} {arXiv:1304.4131 [hep-th]} \BibitemShut
  {NoStop}%
\bibitem [{\citenamefont {Bagnuls}\ and\ \citenamefont
  {Bervillier}(2001)}]{Bagnuls:2000ae}%
  \BibitemOpen
  \bibfield  {author} {\bibinfo {author} {\bibfnamefont {C.}~\bibnamefont
  {Bagnuls}}\ and\ \bibinfo {author} {\bibfnamefont {C.}~\bibnamefont
  {Bervillier}},\ }\href {\doibase 10.1016/S0370-1573(00)00137-X} {\bibfield
  {journal} {\bibinfo  {journal} {Phys.Rept.}\ }\textbf {\bibinfo {volume}
  {348}},\ \bibinfo {pages} {91} (\bibinfo {year} {2001})},\ \Eprint
  {http://arxiv.org/abs/hep-th/0002034} {arXiv:hep-th/0002034 [hep-th]}
  \BibitemShut {NoStop}%
\bibitem [{\citenamefont {Polchinski}(1984)}]{polchinski1984renormalization}%
  \BibitemOpen
  \bibfield  {author} {\bibinfo {author} {\bibfnamefont {J.}~\bibnamefont
  {Polchinski}},\ }\href@noop {} {\bibfield  {journal} {\bibinfo  {journal}
  {Nuclear Physics B}\ }\textbf {\bibinfo {volume} {231}},\ \bibinfo {pages}
  {269} (\bibinfo {year} {1984})}\BibitemShut {NoStop}%
\bibitem [{\citenamefont {Comellas}\ and\ \citenamefont
  {Travesset}(1997)}]{comellas1997n}%
  \BibitemOpen
  \bibfield  {author} {\bibinfo {author} {\bibfnamefont {J.}~\bibnamefont
  {Comellas}}\ and\ \bibinfo {author} {\bibfnamefont {A.}~\bibnamefont
  {Travesset}},\ }\href@noop {} {\bibfield  {journal} {\bibinfo  {journal}
  {Nuclear Physics B}\ }\textbf {\bibinfo {volume} {498}},\ \bibinfo {pages}
  {539} (\bibinfo {year} {1997})}\BibitemShut {NoStop}%
\bibitem [{\citenamefont {Halpern}\ and\ \citenamefont
  {Huang}(1996)}]{Halpern:1995vf}%
  \BibitemOpen
  \bibfield  {author} {\bibinfo {author} {\bibfnamefont {K.}~\bibnamefont
  {Halpern}}\ and\ \bibinfo {author} {\bibfnamefont {K.}~\bibnamefont
  {Huang}},\ }\href {\doibase 10.1103/PhysRevD.53.3252} {\bibfield  {journal}
  {\bibinfo  {journal} {Phys. Rev.}\ }\textbf {\bibinfo {volume} {D53}},\
  \bibinfo {pages} {3252} (\bibinfo {year} {1996})},\ \Eprint
  {http://arxiv.org/abs/hep-th/9510240} {arXiv:hep-th/9510240 [hep-th]}
  \BibitemShut {NoStop}%
\bibitem [{\citenamefont {Periwal}(1996)}]{Periwal:1995hw}%
  \BibitemOpen
  \bibfield  {author} {\bibinfo {author} {\bibfnamefont {V.}~\bibnamefont
  {Periwal}},\ }\href {\doibase 10.1142/S0217732396002885} {\bibfield
  {journal} {\bibinfo  {journal} {Mod. Phys. Lett.}\ }\textbf {\bibinfo
  {volume} {A11}},\ \bibinfo {pages} {2915} (\bibinfo {year} {1996})},\ \Eprint
  {http://arxiv.org/abs/hep-th/9512108} {arXiv:hep-th/9512108 [hep-th]}
  \BibitemShut {NoStop}%
\bibitem [{\citenamefont {{Cheung}}\ \emph {et~al.}(2008)\citenamefont
  {{Cheung}}, \citenamefont {{Fitzpatrick}}, \citenamefont {{Kaplan}},
  \citenamefont {{Senatore}},\ and\ \citenamefont {{Creminelli}}}]{eftinf}%
  \BibitemOpen
  \bibfield  {author} {\bibinfo {author} {\bibfnamefont {C.}~\bibnamefont
  {{Cheung}}}, \bibinfo {author} {\bibfnamefont {A.~L.}\ \bibnamefont
  {{Fitzpatrick}}}, \bibinfo {author} {\bibfnamefont {J.}~\bibnamefont
  {{Kaplan}}}, \bibinfo {author} {\bibfnamefont {L.}~\bibnamefont
  {{Senatore}}}, \ and\ \bibinfo {author} {\bibfnamefont {P.}~\bibnamefont
  {{Creminelli}}},\ }\href {\doibase 10.1088/1126-6708/2008/03/014} {\bibfield
  {journal} {\bibinfo  {journal} {Journal of High Energy Physics}\ }\textbf
  {\bibinfo {volume} {3}},\ \bibinfo {eid} {014} (\bibinfo {year} {2008})},\
  \Eprint {http://arxiv.org/abs/0709.0293} {arXiv:0709.0293 [hep-th]}
  \BibitemShut {NoStop}%
\bibitem [{\citenamefont {{Goldwirth}}\ and\ \citenamefont
  {{Piran}}(1992)}]{goldinitinf}%
  \BibitemOpen
  \bibfield  {author} {\bibinfo {author} {\bibfnamefont {D.~S.}\ \bibnamefont
  {{Goldwirth}}}\ and\ \bibinfo {author} {\bibfnamefont {T.}~\bibnamefont
  {{Piran}}},\ }\href {\doibase 10.1016/0370-1573(92)90073-9} {\bibfield
  {journal} {\bibinfo  {journal} {Phys.Rept.}\ }\textbf {\bibinfo {volume}
  {214}},\ \bibinfo {pages} {223} (\bibinfo {year} {1992})}\BibitemShut
  {NoStop}%
\bibitem [{\citenamefont {{Liddle}}\ and\ \citenamefont
  {{Leach}}(2003)}]{efolds}%
  \BibitemOpen
  \bibfield  {author} {\bibinfo {author} {\bibfnamefont {A.~R.}\ \bibnamefont
  {{Liddle}}}\ and\ \bibinfo {author} {\bibfnamefont {S.~M.}\ \bibnamefont
  {{Leach}}},\ }\href {\doibase 10.1103/PhysRevD.68.103503} {\bibfield
  {journal} {\bibinfo  {journal} {Phys. Rev. D}\ }\textbf {\bibinfo {volume}
  {68}},\ \bibinfo {eid} {103503} (\bibinfo {year} {2003})},\ \Eprint
  {http://arxiv.org/abs/astro-ph/0305263} {astro-ph/0305263} \BibitemShut
  {NoStop}%
\bibitem [{\citenamefont {{Planck Collaboration}}\ \emph
  {et~al.}(2015)\citenamefont {{Planck Collaboration}}, \citenamefont {{Ade}},
  \citenamefont {{Aghanim}}, \citenamefont {{Arnaud}}, \citenamefont
  {{Arroja}}, \citenamefont {{Ashdown}}, \citenamefont {{Aumont}},
  \citenamefont {{Baccigalupi}}, \citenamefont {{Ballardini}}, \citenamefont
  {{Banday}},\ and\ \citenamefont {et~al.}}]{planck2015}%
  \BibitemOpen
  \bibfield  {author} {\bibinfo {author} {\bibnamefont {{Planck
  Collaboration}}}, \bibinfo {author} {\bibfnamefont {P.~A.~R.}\ \bibnamefont
  {{Ade}}}, \bibinfo {author} {\bibfnamefont {N.}~\bibnamefont {{Aghanim}}},
  \bibinfo {author} {\bibfnamefont {M.}~\bibnamefont {{Arnaud}}}, \bibinfo
  {author} {\bibfnamefont {F.}~\bibnamefont {{Arroja}}}, \bibinfo {author}
  {\bibfnamefont {M.}~\bibnamefont {{Ashdown}}}, \bibinfo {author}
  {\bibfnamefont {J.}~\bibnamefont {{Aumont}}}, \bibinfo {author}
  {\bibfnamefont {C.}~\bibnamefont {{Baccigalupi}}}, \bibinfo {author}
  {\bibfnamefont {M.}~\bibnamefont {{Ballardini}}}, \bibinfo {author}
  {\bibfnamefont {A.~J.}\ \bibnamefont {{Banday}}}, \ and\ \bibinfo {author}
  {\bibnamefont {et~al.}},\ }\href@noop {} {\bibfield  {journal} {\bibinfo
  {journal} {ArXiv e-prints}\ } (\bibinfo {year} {2015})},\ \Eprint
  {http://arxiv.org/abs/1502.02114} {arXiv:1502.02114} \BibitemShut {NoStop}%
\bibitem [{\citenamefont {Szeg{\H{o}}}(1939)}]{szego1939orthogonal}%
  \BibitemOpen
  \bibfield  {author} {\bibinfo {author} {\bibfnamefont {G.}~\bibnamefont
  {Szeg{\H{o}}}},\ }\href {https://books.google.nl/books?id=ZOhmnsXlcY0C}
  {\emph {\bibinfo {title} {Orthogonal Polynomials}}},\ \bibinfo {series}
  {American Mathematical Society colloquium publications}\ No.\ \bibinfo
  {number} {v. 23, pt. 2}\ (\bibinfo  {publisher} {American Mathematical
  Society},\ \bibinfo {year} {1939})\BibitemShut {NoStop}%
\end{thebibliography}%

\end{document}